\documentclass[aps,superscriptaddress,showpacs]{revtex4}

\usepackage{amsmath}
\usepackage{graphicx}

\begin{document}

\title{Nonlinear aspects of quantum plasma physics}
\author{P K Shukla}
\affiliation{Institut f\"ur Theoretische Physik IV, Fakult\"at f\"ur Physik und Astronomie, Ruhr--Universit\"at Bochum, D-44780 Bochum, Germany}
\affiliation{Scottish Universities Physics Alliance (SUPA, Department of
Physics, University of Strathclyde, Glasgow G4 0NG, United Kingdom}
\affiliation{Instituto de Plasmas e Fus\~ao Nuclear, Instituto Superior T\'ecnico, Universidade T\'ecnica de Lisboa, 1049-001 Lisboa, Portugal}
\affiliation{Department of Physics, Ume\aa~University, SE-90 187 Ume\aa,~Sweden}
\affiliation{The Abdus Salam International Centre for Theoretical Physics, I-34014 Trieste, Italy}
\author{B Eliasson}
\affiliation{Institut f\"ur Theoretische Physik IV, Fakult\"at f\"ur Physik und Astronomie, Ruhr--Universit\"at Bochum, D-44780 Bochum, Germany}
\affiliation{Department of Physics, Ume\aa~University, SE-90 187 Ume\aa,~Sweden}
\affiliation{The Abdus Salam International Centre for Theoretical Physics, I-34014 Trieste, Italy}

\received{25 June 2009}
\revised{26 August 2009}

\begin{abstract}
Dense quantum plasmas are ubiquitous in planetary interiors and in compact astrophysical objects
(e.g. the interior of white dwarf stars, in magnetars etc.), in semiconductors and micro-mechanical
systems, as well as in the next generation intense laser-solid density plasma interaction experiments
and in quantum x-ray free-electron lasers. In contrast to classical plasmas, one encounters extremely
high plasma number density and low temperature in quantum plasmas. The latter are composed of electrons,
positrons and holes, which are degenerate. Positrons (holes) have the same (slightly different) mass as
electrons, but opposite charge. The degenerate charged particles (electrons, positrons, holes) follow the
Fermi-Dirac statistics.  In quantum plasmas, there are new forces associated with i) quantum statistical
electron and positron pressures, ii) electron and positron tunneling through the Bohm potential,
and iii) electron and positron angular momentum spin. Inclusion of these quantum forces provides possibility
of very high-frequency dispersive electrostatic and electromagnetic waves (e.g. in the hard x-ray and gamma
rays regimes) having extremely short wavelengths. In this review paper, we present theoretical backgrounds
for some important nonlinear aspects of wave-wave and wave-electron interactions in dense quantum plasmas.
Specifically, we shall focus on nonlinear electrostatic electron and ion plasma waves, novel aspects of
3D quantum electron fluid turbulence, as well as nonlinearly coupled intense electromagnetic waves
and localized plasma wave structures. Also discussed are the phase space kinetic structures and
mechanisms that can generate quasi-stationary magnetic fields in dense quantum plasmas. The influence of
the external magnetic field and the electron angular momentum spin on the electromagnetic wave dynamics
is discussed. Finally, future perspectives of the nonlinear quantum plasma physics are highlighted.
\end{abstract}

\maketitle

\tableofcontents

\newpage

\section{Introduction}

The field of quantum plasma physics has a long and diverse tradition
\cite{Klimontovich52,Bohm52,Bohm53,Pines1,Pines2}, and is becoming of increasing current
interest\cite{Bonitz03,Manfredi05}, motivated by its potential applications in modern technology
(e.g. metallic and semiconductor nanostructures-such as metallic nanoparticles, metal clusters,
thin metal films, spintronics, nanotubes, quantum well and quantum dots, nano-plasmonic devices,
quantum x-ray free-electron lasers, etc.). Due to the recent development of ultrafast spectroscopy techniques,
it is now possible to monitor the femtosecond dynamics of an electron gas confined in metallic plasmas.
In dense quantum plasmas, the number densities of degenerate electrons and/or positrons are extremely
high, and the plasma particles (mainly electrons and positrons) obey Fermi-Dirac statistics.

The quantum degeneracy effects start playing a significant role when the de~Broglie thermal wavelength
$\lambda_B =\hbar/(2\pi m_e k_B T)^{1/2}$ for electrons is similar to or larger than the average
inter-electron distance $n_e^{-1/3}$, i.e. when \cite{Bonitz03,Manfredi05}
\begin{equation}
  n_e \lambda_B^3 \gtrsim 1,
\end{equation}
or, equivalently, the temperature $T$ is comparable or lower than the electron Fermi temperature
$T_{Fe}=E_F/k_B$, where the electron Fermi energy is
\begin{equation}
E_F =\frac{\hbar^2}{2m_e} (3\pi^2)^{2/3}n_e^{2/3},
\end{equation}
so that
\begin{equation}
\chi =\frac{T_{Fe}}{T} = \frac{1}{2}(3\pi^2)^{2/3}({n_e\lambda_B^3})^{2/3}\gtrsim 1.
\end{equation}
Here $\hbar$ is the Planck constant divided by $2\pi$, $n_e$ is the electron number density,
$m_e$ is the rest electron mass,  and $k_B$ is the Boltzmann constant.

When the temperature approaches the electron Fermi temperature $T_{Fe}$, one can show, by using the
density matrix formalism \cite{Bransden00}, that the equilibrium electron distribution function changes
from the Maxwell--Boltzmann $\propto \exp(-E/k_B T)$ to the Fermi--Dirac distribution
$\propto (2/\hbar^3)\left[\exp((E+\mu)/k_B T_{Fe})+1\right]^{-1}$, where $E$ is the electron kinetic energy
and $\mu$ is the chemical potential. In a dense Fermi plasma, the Thomas-Fermi screening radius reads
\begin{equation}
\lambda_F =\frac{V_{Fe}}{\sqrt{3}\omega_{pe}}
\end{equation}
which is the quantum analogue of the Debye-H\"uckel radius. Here the electron Fermi speed
\begin{equation}
 V_{Fe} =(2 E_F/m_e)^{1/2} = \frac{\hbar}{m_e}(3\pi^2 n_e)^{1/3}
\end{equation}
is the speed of an electron at the Fermi surface.

\begin{figure}[htb]
\includegraphics[width=10cm]{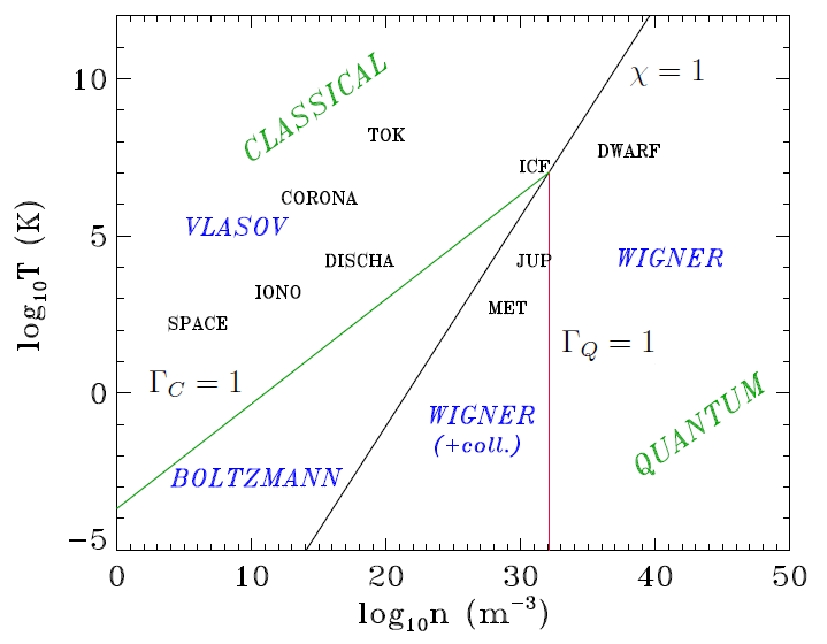}
\caption{Schematic plasma diagram in the $\log_{10}(T)$ -- $\log(n)$ plane: IONO: ionospheric plasma, SPACE:
space plasma, CORONA: solar corona, DISCHA: typical electric discharge, TOK: tokamak/magnetic fusion experiments,
ICF: inertial confinement fusion, MET: metals and metal clusters, JUP: Jupiter's core, DWARF: white dwarf star.
After Refs. \cite{Bonitz03,Manfredi05}.}
\label{Fig:Plasma_diagram}
\end{figure}

A measure of the importance of collisions in a dense plasma is the quantum coupling parameter, which is the
ratio between the interaction energy $E_{int}=e^2 n_e^{1/3}$ and the average kinetic energy $E_{kin}$ of
electrons, where $e$ is the magnitude of the electron charge.
For a classical plasma, the kinetic energy is $k_B T$ and hence we have $\Gamma_C ={E_{int}}/{k_B T}$ in the
classical case. In a quantum plasma, we have instead $E_{kin}=E_F$, which gives the quantum coupling
parameter \cite{Bonitz03,Manfredi05,Fortov00}
\begin{equation}
\Gamma_Q =\frac{E_{int}}{E_F}=\frac{2}{(3\pi^2)^{2/3}}\frac{m_e e^2}{\hbar^2 n_e^{1/3}} \sim
\left(\frac{1}{n_e\lambda_F^3}\right)^{2/3}
\sim \left(\frac{\hbar \omega_{pe}}{2k_B T_{Fe}}\right)^2 \equiv H^2,
\end{equation}
(where we have left out proportionality constants for the sake of clarity) is analogous to the classical
one when $\lambda_F \rightarrow \lambda_D$. The various plasma regimes are
illustrated in Fig. \ref{Fig:Plasma_diagram}, where the straight lines correspond to
i) the distinction between the classical and quantum plasmas, $\chi=1$, ii) the limit between collisionless
and collisional classical plasmas, $\Gamma_C=1$, and iii) the limit between collisionless and collisional
quantum plasmas, $\Gamma_Q=1$. In Fig. \ref{Fig:Plasma_diagram}, various experimental and naturally occurring
plasmas are exemplified. Also indicated in Fig. \ref{Fig:Plasma_diagram} are the kinetic equations used to model
the plasma in each regime. The Vlasov and Wigner equations are thus used to model collisionless plasma in the classical and
quantum limits, respectively, while "Boltzmann" indicates collisional kinetic models in a classical plasma.
The kinetic models for a collisional quantum plasma is labeled "Wigner (+coll)" in Fig. \ref{Fig:Plasma_diagram}.
We note that a quantum plasma becomes collisionless $\Gamma_Q < 1$ when the mean distance between electrons is of
the order of the Bohr radius $a_0$, i.e. $d=1/n_e^{1/3}<[(3\pi^2)^{2/3}/8\pi]a_0\approx 0.38 a_0$, where
$a_0=\hbar^2/m_e e^2 \approx 0.53$ \AA. This corresponds to a number density of the order $n_e \gtrsim 1.22\times 10^{26}\,\mathrm{cm^{-3}}$, which is three orders of magnitude larger
than the electron density in a typical metal, for example gold has $n_e =5.9\times10^{22}\,\mathrm{cm^{-3}}$
at room temperature.

In Fermi-degenerate matters, there is an effect called the Pauli-blocking, which strongly reduces the
electron-electron and electron-ion collisional rates. Namely, at moderate temperatures, only electrons
within an energy shell of thickness $k_B T$ about the Fermi surface (where the electron energy equals $E_F$)
can undergo collisions. For these electrons, the electron-electron collision rate is of the order
$k_B T/\hbar$, and the average collision rate is obtained by multiplying this expression by $T/T_F$.
The resulting collision frequency $\nu_{ee}$ divided by the electron plasma frequency is \cite{Manfredi05}
\begin{equation}
\frac{\nu_{ee}}{\omega_{pe}}\sim \frac{E_F}{\hbar \omega_p}\left(\frac{T}{T_F}\right)^2
=\frac{1}{\Gamma_Q^{1/2}}\left(\frac{T}{T_F}\right)^2.
\end{equation}
Hence, $\nu_{ee}\ll\omega_{pe}$ when $T\ll T_F$ and $\Gamma_Q>1$, which is relevant for metallic electrons.
For example, at room temperature we have $\nu_{ee}\sim 10^{11}\,s^{-1}$, which is much smaller than the
typical collisionless frequency of collective interactions, $\omega_{pe}\sim 10^{16}\,\mathrm{s}^{-1}$.
Also, the typical collision rate for electron-lattice (ion) collisions, $\nu_{ei}\sim 10^{15}\,\mathrm{s}^{-1}$,
is smaller than $\omega_{pe}$ by one order of magnitude. Therefore, a collisionless regime seems to be
relevant for free electrons in a metal on a time-scale of the order of a femtosecond. In denser plasmas,
such as in stellar interiors \cite{Lee50,Hubbard66,Lampe68} and in inertial fusion schemes \cite{Azechi91,Azechi06,Son05}, the relative effects of collisions decrease even further and leads to increased electron transport and heat conductivity.

More than sixty five years ago, Wigner \cite{Wigner32} introduced a phase-space formalism to treat a quantum
state of charged particles in a collisionless quantum system. He introduced a quantum distribution function of
the phase-space variables $f(x,p.t)$.  The Wigner distribution function is defined as
\begin{equation}
f(x,p,t) =\frac{1}{(2\pi)^{3N}}\int_\infty^\infty
\rho\bigg(x-\frac{\hbar}{2}\tau, x+ \frac{\hbar}{2} \tau, t\bigg)
\exp(-ip \tau) d \tau,
\end{equation}
where $N$ is the number of particles in the system, $x\{x;1,x_2..., x_N\}$ and $p\{p_1,p_2,...,P_N\}$ are
the sets of coordinates and momenta for the particles, $t$ is the time,
and $\rho(x,x^{\prime},t)$ is the density matrix. The Wigner function is not a
probability density in phase space $x,p$, because it can take negative values.  The Wigner distribution
function $W_i(x,p.t)$, corresponding to the wave function $\psi_i(x,t)$, can be expressed as
\begin{equation}
W_i(x,p, t)=\frac{1}{(2\pi)^{3N}\hbar}\int_\infty^\infty dy
\left<\psi_i(x-y/2,t)\psi_i^*(x+y/2,t)\right>\exp(-py/\hbar),
\end{equation}
which has the property
\begin{equation}
\int_\infty^\infty dp W_i(x,p.t) = <|\psi_i(x,t)|^2>,
\end{equation}
where the asterisk denotes complex conjugate.  The  quantum kinetic equation based on the Wigner
distribution was developed by Moyal \cite{Moyal49}, and is now referred to as the Wigner-Moyal
description for treating statistical effects on electron plasma waves in a quantum plasma \cite{Anderson02}.

Analytical investigations of collective interactions between an ensemble of degenerate electrons in a dense
quantum plasma dates back to early fifties. Specifically, Klimontovich and Silin \cite{Klimontovich52} and
Bohm and Pines \cite{Bohm52,Bohm53,Pines1,Pines2} presented properties of linear electron plasma oscillations (EPOs)
in a dense quantum plasma.  In  the latter, electrons, positrons, and holes are degenerate, while ions
are cold and classical (typically, in a dense quantum plasma the ion Fermi speed is much smaller
than the electron Fermi speed). Accordingly, electrons, positrons and holes have a Fermi-Dirac
distribution function \cite{Tsintsadze09}, contrary to the Boltzmann-Maxwell distribution function for charged particles in
a classical plasma.

The dispersion relation for high-frequency electron plasma oscillations in a dense quantum plasma with fixed
ion background reads (see Appendix B)
\begin{equation}
  \begin{split}
  &1-\frac{4 \pi e^2}{m_e}
   \int
     \frac{f_0({\bf u})}{(\omega-{\bf k}\cdot{\bf u})^2-\frac{\hbar^2 k^4}{4 m_e^2}} d^3 u=0,
  \end{split}
\end{equation}
which was also obtained by Bohm and Pines \cite{Bohm53} by performing a series of canonical transformations on
the Hamiltonian of the system of individual electrons, interacting via the electrostatic force. In the zero temperature
limit, we have (see Appendix B)
\begin{equation}
  \begin{split}
  1+\frac{3\omega_{pe}^2}{4 k^2 V_{Fe}^2}\left\{
  2-\frac{m_e}{\hbar k V_{Fe}}\left[V_{Fe}^2-\left(\frac{\omega}{k}+\frac{\hbar k}{2 m_e}\right)^2\right]
  \log\left|
  \frac{
  \frac{\omega}{k}-V_{Fe}+\frac{\hbar k}{2 m_e}
  }{
  \frac{\omega}{k}+V_{Fe}+\frac{\hbar k}{2 m_e}
  }
  \right|
  \right.
  \\
  \left.
  +\frac{m_e}{\hbar k V_{Fe}}\left[V_{Fe}^2-\left(\frac{\omega}{k}-\frac{\hbar k}{2 m_e}\right)^2\right]
  \log\left|
  \frac{
  \frac{\omega}{k}-V_{Fe}-\frac{\hbar k}{2 m_e}
  }{
  \frac{\omega}{k}+V_{Fe}-\frac{\hbar k}{2 m_e}
  }
  \right|
  \right\}\equiv 1+\chi_e=0,
  \end{split}
  \label{Wigner_disp}
\end{equation}
which, in the limit $\hbar k/m_e\rightarrow 0$ yields
\begin{equation}
  1+\frac{3\omega_{pe}^2}{k^2 V_{Fe}^2}\left(
  1-\frac{\omega}{2k V_{Fe}}\log\left|\frac{\omega+k V_{Fe}}{\omega-k V_{Fe}}\right|
  \right)=0,
\end{equation}
where we have assumed that $\omega$ is real and $\omega/k>V_{Fe}$.
Here $\omega$ is the wave frequency, ${\bf k}$ is the wave vector, and $\omega_{pe} =(4\pi n_e e^2/m_e)^{1/2}$
is the electron plasma frequency,
On the other hand,
for small wavenumbers up to terms containing $k^4$, we obtain from (\ref{Wigner_disp})
\begin{equation}
  \omega^2\approx\omega_{pe}^2+\frac{3}{5}k^2 V_{Fe}^2+(1+\alpha)\frac{\hbar^2 k^4}{4 m_e^2},
\end{equation}
where $\alpha=(48/175)m_e^2 V_{Fe}^4/\hbar^2\omega_{pe}^2\approx 2.000 (a_0^3 n_0)^{1/3}$, where $a_0=\hbar^2/m_e e^2\approx 53\times 10^{-10}\,\mathrm{cm}$ is the Bohr radius. Equation (14) shows that the wave dispersion arises due to the finite width of
the electron wave function in  a dense Fermi
plasma \cite{Gardner96,Manfredi01,Manfredi05,Shukla06,Shukla06a,ShuklaEliasson07}.

Furthermore, in the low phase speed limit, viz. $\omega \ll k V_{Fe}$, the dielectric
constant for ion oscillations reads
\begin{equation}
\epsilon(\omega, {\bf k}) \approx 1 + \frac{3\omega_{pe}^2}{k^2 V_{Fe}^2
+ 3\hbar^2k^4/4m_e^2}-\frac{\omega_{pi}^2}{\omega^2},
\end{equation}
which yields, with $\epsilon(\omega, {\bf k}) =0$, the ion oscillation frequency
\begin{subequations}
\begin{equation}
\omega \approx \frac{\omega_{pi}}{(1+Q)^{1/2}},
\end{equation}
where $\omega_{pi} =(m_e/m_i)^{1/2}\omega_{pe}$
is the ion plasma frequency, and $m_i$ is the ion mass, $ Q = 3\omega_{pe}^2/(k^2 V_{Fe}^2+ 3\hbar^2 k^4/4m_e^2)$, and $\alpha\ll 1$.
For $Q \gg 1$, we have from (16b)
\begin{equation}
\omega \approx kC_{Fs}  \left(1+ \frac{\hbar^2k^4}{4m_e^2\omega_{pe}^2}\right)^{1/2},
\end{equation}
\end{subequations}
where $C_{Fs} =(T_{Fe}/3 m_i)^{1/2}$ is the sound speed.

Dispersion properties of electrostatic waves in an unmagnetized dense quantum plasmas with
arbitrary electron degeneracy have been presented by Maafa \cite{Maafa93} and Melrose \cite{Melrose07}.
By using the random phase approximation, the permittivity of a degenerate collisionless plasma is
given in textbooks \cite{Lifshitz81,Nozieres99}. Furthermore, since the pure electromagnetic wave in
a non-streaming unmagnetized dense plasma do not accompany density fluctuations, the wave frequency
is $\omega=(\omega_{pe}^2+k^2c^2)^{1/2}$, where $c$ is the speed of light in vacuum. Theoretical studies
of quantum statistical properties of dense plasmas in the presence of electromagnetic waves appear
in Kremp et al. \cite{Kremp99} and in textbooks \cite{Bonitz98,Kremp05}, while quantum
parameter regimes are discussed by Bonitz \cite{Bonitz03}. In a magnetized dense
quantum plasma, one finds that the external magnetic field  significantly affects the dynamics
of degenerate electrons and positrons, and subsequently there appear new collective phenomena
associated with the electron angular momentum spin \cite{Uhlenbeck25,Dbohm52}, the electron spin
magnetic moment \cite{Sasabe08}, and quantized Landau energy levels \cite{Landau77} of
the Fermions in a strong magnetic field. It turns out that the thermodynamics and kinetics \cite{Steinberg00},
as well as the dispersion properties of both electrostatic and electromagnetic
waves \cite{John06,Shukla06b,ShuklaAli06,Shukla07a,Lundin07,Misra07,Saleem09}
in a quantum magnetoplasma are significantly different from those in an unmagnetized quantum plasma.

It was early recognized that the underlying physics of nonlinear quantum-like equations can be
better understood by casting those equations in the form of hydrodynamical (or the Euler)
equations, which essentially represent the evolution of quantum particle densities and momenta.
This was elegantly done by Madelung \cite{Madelung26} and Bohm \cite{Bohm52} by introducing an
eikonal representation for the wave function evolution in the non-stationary Schr\"odinger equation.
The derivation of the Madelung quantum fluid equations for the Pauli equation with the quantum particle
angular momentum spin was presented by Takabayashi \cite{Takabayashi52,Takabayashi55a,Takabayashi83},
Bohm et al. \cite{Bohm55a,Bohm55b}, Janossy and Ziegler-Naray \cite{Janossy65}, and others
\cite{Ghosh82}. To incorporate relativistic effects into the quantum fluid formalism, Takabayashi
derived the quantum electron fluid equations for the Klein-Gordon equation \cite{Takabayashi53}
and for the Dirac equation \cite{Takabayashi55b,Takabayashi56,Takabayashi57}. Extensions have also been
done to fluid descriptions of the Weyl equation for massless spin-1/2 particles (neutrinos) by
Bialynicki-Birula \cite{Bialy95}.

Recently, there has been growing and vibrant interests in investigating new aspects of quantum
plasma physics by developing non-relativistic quantum hydrodynamical (QHD) equations
\cite{Gardner94,Gardner96,Manfredi01,Manfredi05}. The latter include the quantum statistical
electron pressure and the quantum force involving tunneling of degenerate electrons through
the Bohm potential \cite{Gardner96}. The Wigner-Poisson (WP) model has also been used to derive
a set of non-relativistic quantum hydrodynamical (QHD) equations \cite{Manfredi01,Manfredi05} for
a dense electron plasma, assuming immobile ions. The QHD equations are composed of the electron
continuity, electron momentum and Poisson equations. The quantum force \cite{Gardner96,Manfredi01,Manfredi05}
appears in the non-relativistic electron momentum equation through the pressure term, which requires
knowledge of the Wigner distribution for a quantum mixture of electron wave functions, each characterized
by an occupation probability. Quantum transport models similar to the QHD plasma model have also been
used in superfluidity \cite{r9} and superconductivity \cite{r10}, as well as in the study of metal
clusters and nanoparticles, where they are referred to as nonstationary Thomas-Fermi models \cite{r11}.

The electrostatic QHD equations are useful for studying collective interactions (e.g. different types
of waves, instabilities, quantum fluid turbulence and nonlinear structures
\cite{Manfredi01,Manfredi05,Haas03,Eliasson08,Haas05,Shukla06,ShuklaStenflo06,Shukla07,Shaikh07,
Shaikh08,Bersh08,Eliasson08a} in dense quantum plasmas. The quantum kinetic and QHD equations have also been
generalized to include the electromagnetic, ambient magnetic field and electron angular momentum spin
effects \cite{Semikoz03,MarklundBrodin07,Brodin07,Shukla07a,Shukla08b,Shukla09a}.
The latter give rise to high-frequency spin waves, which can be excited by neutrino beams in
supernovae \cite{Semikoz02a,Semikoz03}. Furthermore, studies of numerous collective interactions
in dense plasmas are relevant in the context of i) intense laser-solid density plasma experiments
\cite{Hu99,Andreev00,Mourou06,MarklundShukla06,Salamin06,Azechi91,Azechi06,Malkin07,Kritcher08,Hartemann08,Lee09,Norreys09},
where one would be exploring new frontiers in high-energy density physics \cite{Drake09};
ii)  in the cores of giant planets and the crusts of old stars \cite{Horn91,Guillot99,Fortney09};
iii) superdense astrophysical objects \cite{Meszaros92,Gurevich93,Craighead00,Opher01,Shapiro04,Benvenuto05,
Chabrier02, Chabrier06,Lai06} (e.g. interiors of white dwarfs and magnetospheres of neutron stars and magnetars);
iv) as well as for micro and nano-scale objects (e.g. quantum diodes \cite{Lao91,Ang03,Ang04,Ang06,Ang07,Shukla08},
quantum dots and nanowires \cite{r13D}, nano-photonics \cite{r14D,r14Db}, plasmonics \cite{r14Dc},
ultra-small electronic devices \cite{r0,r1,r1m}, and metallic nanostructures \cite{r1a});
v) micro-plasmas \cite{r2}, and quantum x-ray free-electron lasers \cite{r3,r4}. Furthermore,
it should be stressed that a Fermi degenerate dense plasma may also arise when a pellet of hydrogen
is compressed to many times the solid density in the fast ignition scenario for inertial confinement
fusion \cite{Son05,Lindl95,Tabak05}.  Since there is an impressive developments in the field of short pulse
petawatt laser technology, it is highly likely that such plasma conditions can be achieved by intense laser pulse
compression using  powerful x-ray pulses. Here ultrafast x-ray Thomson scattering techniques can be used to
measure the features of laser enhanced plasma lines, which will, in turn, give invaluable informations
regarding the equation of state of shock compressed dense matters. Recently, spectrally resolved x-ray scattering
measurements \cite{Kritcher08,Lee09} have been performed in dense plasmas allowing accurate measurements of
the electron velocity distribution function, temperature, ionization state, and of plasmons in the
warm dense matter regime \cite{Glenzer07}. This novel technique promises to access the degenerate, the closely
coupled, and the ideal plasma regime, making it possible to investigate extremely dense states of matter,
such as the inertial confinement fusion fuel during compression, reaching super-solid densities.

In this review article, we present the theoretical progress that has been recently made in the area of
collective nonlinear interactions in collisionless dense quantum plasmas. The manuscript is organized in
the following fashion. In section 2, we shall briefly recapitulate the hydrodynamic representation of
some quantum-like models that appear in different branches of physics. The governing equations for
nonlinearly interacting electrostatic waves in an unmagnetized quantum  plasma are derived in
section 3. Section 4 presents numerical studies of nonlinear electron plasma wave excitations
in the form of quantized one-dimensional dark solitons and quantized two-dimensional vortices.
The model used here is the nonlinear Schr\"odinger equation for the dispersive EPOs, coupled with
the Poisson equation for the electrostatic potential. This model is also used for studying 3D
quantum electron fluid turbulence in section \ref{sec:turbulence}, where we find non-Kolmogorov-type
turbulence spectra. In section \ref{sec:kinetic}, we present recent results concerning the phase
space (kinetic) turbulence, by using the Wigner and Vlasov models for the electron distribution function.
A theoretical model for the generation of quasi-stationary magnetic fields in a dense quantum plasma due
to the Weibel instability is presented in section \ref{sec:Weibel}. The magnetization of a dense plasma
in the presence of a large amplitude electromagnetic wave is demonstrated in subsection \ref{sec:Magnetization}.
The dynamics of electromagnetic waves in a dense magnetoplasma are discussed in section \ref{emwaveequation}.
Here we focus on spin waves propagating across the magnetic field direction, and
develop nonlinear equations for low-phase speed (in comparison with the speed of light)
electromagnetic waves in a dense quantum  magnetoplasma. Finally, section \ref{sec:conclusions}
highlights our main results and describes the future prospectives of the quantum plasma
physics research.

\section{Fluid representation of quantum-like models}

This section is included to show how different types of quantum-like models can be
cast in the form of hydrodynamic equations.

First, we consider the non-stationary nonlinear Schr\"odinger equation(NLSE)
\begin{equation}
i \hbar\frac{\partial \psi}{\partial t}+\frac{\hbar^2}{2 m}\nabla^2\psi - U_{0}(|\psi|^2) \psi =0,
\label{Schrod}
\end{equation}
where $\psi({\bf r}, t)$ is the macroscopic wave function, $m$ is the particle mass, and $U_0(|\psi|^2)$
is an effective potential. The NLSE also arises in various physical context in the description of
amplitude modulated nonlinear waves in fluids \cite{Benney67,Benjamin67}, in transmission
lines \cite{Scott73}, in nonlinear optics for ultra-fast communications \cite{Hasegawa73,Agrawal06,Wan07},
in plasmas \cite{Karpman69,Karpman71,Zakharov72,Karpman75,SchamelShukla76,Shukla78}, and in many other
areas of physics \cite{Sulem99,Fedele02,Dauxois06}.

Introduce the Madelung transformation \cite{Madelung26}
\begin{equation}
\psi ({\bf r}, t) =\sqrt{n}\exp \left(i \frac{\varphi_q}{\hbar} \right),
\end{equation}
where $n$ and $\varphi_q$ are real, and obtain from (17) a pair of quantum hydrodynamic equations
composed of the continuity and momentum equations, respectively,
\begin{equation}
\frac{\partial n}{\partial t}+\nabla\cdot(n{\bf v})=0,
\label{cont}
\end{equation}
and
\begin{equation}
m\left(\frac{\partial}{\partial t}+{\bf v}\cdot \nabla \right){\bf v} = -\nabla [U_0(n) + U_B].
  \label{mom}
\end{equation}
Here $n = n({\bf r}, t) = |\psi|^2$ corresponds to the local density per unit length,
and $ \hbar \nabla \varphi_q({\bf r}, t) =m {\bf v}$. The quantum potential is
\begin{equation}
U_B=-\frac{\hbar^2}{2 m}\frac{\nabla^2\sqrt{n}}{\sqrt{n}}.
\end{equation}

We note that the quantum particle number density $n$ and the quantum velocity field ${\bf v}$
can be written as, respectively,
\begin{equation}
n ({\bf r}, t) = \psi \psi^* \equiv |\psi|^2,
\end{equation}
and
\begin{equation}
{\bf v}=\frac{\hbar}{2 i m}\frac{(\psi^*\nabla\psi-\psi\nabla\psi^*)}{|\psi|^2}
=-\frac{i\hbar}{2m}\nabla \left[{\rm ln} \left(\frac{\psi}{\psi^*}\right)\right].
\end{equation}
where the asterisk denotes the complex conjugate. The quantum-like velocity, given by (23),
is a potential field, namely,

\begin{equation}
\nabla \times {\bf v} =0
\end{equation}
everywhere in a single-connected region.

Let us now define the generalized vorticity on the weighted velocity field as \cite{Bersh08}
\begin{equation}
\boldsymbol{\Omega} = \frac{\nabla \times(|\psi|^2{\bf v})}{|\psi|^2}
= \nabla \times {\bf v} + \frac{\nabla |\psi|^2 \times {\bf v }}{|\psi|^2},
\end{equation}
where the first term in the right-hand side in (25) represents the ordinary vorticity. It is well
known \cite{Ghosh82,Barenghi01} that in the condensate state all rotational flow is carried by quantized
vortices (the circulation of the velocity around the core of each such vortex is quantized). In the absence
of quantized vortices, the first term is zero in view of (24). In such a situation, the second term in
(25) determines the generalized vorticity. Various aspects of quantized vortex dynamics and superfluid
turbulence appear in Barenghi et al \cite{Barenghi01}. Bewley et al \cite{Sreeni06} have presented
a technique for visualization of quantized vortices in liquid helium.

Second, the nonlinear Schr\"odinger equation can be generalized by including the trapping potential

\begin{equation}
V_b (x,y,z) = \frac{1}{2} m_b \left(\omega_x^2 x^2 + \omega_y^2 y^2 + \omega_z^2 z^2\right),
\end{equation}
which confines identical bosons \cite{Bose24} in the harmonic trap of an ultracold quantum system.
Here  $m_b$ is the boson mass, and $\omega_x$, $\omega_y$ and $\omega_z$ are the harmonic
frequencies of the bosons along the $x, y,$ and $z$ directions, respectively, of a
Cartesian coordinate system.  The nonlinear dynamics of the Bose-Einstein condensates (BECs)
\cite{Bose24,Einstein24} is then governed by the Gross-Pitaevskii equation \cite{Gross61,Pita61,Pita03}

\begin{equation}
i\hbar \frac{\partial \phi ({\bf r},t)}{\partial t} + \frac{\hbar^2}{2m} \nabla^2 \psi ({\bf r},t)
- V_b \psi ({\bf r},t) - G |\psi|^2({\bf r}, t) \psi ({\bf r}, t) =0,
\end{equation}
where for the BECs we have  $U_0(|\psi|^2) =(4\pi \hbar^2a_s/m_b)^{1/2}|\psi|^2 \equiv G|\psi|^2$.
Here $m_b$ is the mass of the bosons and $a_s$ is the scattering length for boson-boson collisions.
The BECs are repulsive (attractive) for $G >$ $(< ) 0$.

Introducing the Ansatz $\psi ({\bf r},t) =\sqrt{n_b({\bf r}, t)} \exp[\varphi_b({\bf r}, t)]$ in (27),
we obtain generalized quantum hydrodynamic equations \cite{Dalfovo99,Dell04}
\begin{equation}
\frac{\partial n_b}{\partial t} + \nabla \cdot (n {\bf u}_b) =0,
\end{equation}
and
\begin{equation}
m_b \frac{\partial {\bf u}_b}{\partial t}
= - \nabla \left(V_b + + \frac{m_b}{2} u_b^2 + G n_b
- \frac{\hbar^2}{2m_b\sqrt{n_b}} \nabla^2\sqrt{n_b}\right),
\end{equation}
where the particle flux
\begin{equation}
n_b({\bf r},t) {\bf u}_b ({\bf r}, t) = \frac{\hbar^2}{2i m_b}
\left(\psi^* \nabla \psi-\psi \nabla \psi^*\right),
\end{equation}
with ${\bf u}_b =(\hbar/m_b)\nabla \varphi_b({\bf r}, t)$. Equation (29) establishes
the irrotational nature of the superfluid motion of the BECs. Equations (29) and (30)
can be used to study the linear and nonlinear properties of BECs.

Next, we consider the dynamics of a non-relativistic single Fermi ($1/2-$ spin) particle
(a degenerate electron) governed by the Pauli equation \cite{Pauli25,Bere99}
\begin{equation}
 i\hbar\frac{\partial \Psi}{\partial t} + \frac{\hbar^2}{2 m_e} \nabla^2 \Psi
-\left[\frac{i e\hbar}{2m_e c} \left({\bf A} \cdot \nabla  + \nabla \cdot {\bf A}\right)
+ \frac{e^2{\bf A}^2}{2m_ec^2} - e \phi - \mu_e {\bf \sigma} \cdot{\bf B}\right]\Psi =0,
  \label{Schrod2}
\end{equation}
where $\Psi ({\bf r}, t, {\boldsymbol \sigma})$ is the wave function of the single particle
species having the spin ${\bf s} =1/2{\boldsymbol \sigma}$, ${\boldsymbol \sigma}$ is the Pauli
spin matrices, ${\bf A}$ is the vector potential, $\phi$ is the scalar potential,
${\bf B} = \nabla \times {\bf A}$, and $\mu_B=e\hbar/2m_ec$ is the Bohr magneton.

By using the Madelung representation for the complex wave function \cite{Brodin07R}
\begin{equation}
\Psi ({\bf r}, t, {\boldsymbol \sigma})
= s \sqrt{n_e({\bf r}, t, {\boldsymbol \sigma})} \exp\left[\frac{i S_e ({\bf r}, t,
{\boldsymbol \sigma})}{\hbar}\right],
\end{equation}
one can obtain from (31) the quantum magnetohydrodynamic equation \cite{Brodin07R,Tsintsadze09}
\begin{equation}
\frac{\partial n_e}{\partial t} + \nabla \cdot \left( \frac{n {\bf p}_e}{m_e}\right) =0,
\end{equation}
and
\begin{equation}
\left(\frac{\partial}{\partial t} + \frac{1}{m_e} {\bf p}_e \cdot\nabla \right) {\bf p}_e
= e \left[\nabla \phi +\frac{1}{c}\frac{\partial {\bf A}}{\partial t}
- \frac{1}{c} {\bf v}_e \times (\nabla \times {\bf A}) \right]
+ \frac{\hbar^2}{2m_e} \nabla\left(\frac{\nabla^2 \sqrt{n_e}}{\sqrt{n_e}}\right)
+ \mu_B \nabla ({\boldsymbol \sigma} \cdot {\bf B}),
\end{equation}
where $s$ mimics the spinor through which the electron-$1/2$ spin properties are mediated,
$n_e ({\bf r}, t, {\boldsymbol \sigma})= \Psi \Psi^*$ represents the probability density
of finding the single electron at some point with a spin ${\bf s}$. We have denoted the generalized electron momentum ${\bf p}_e =\nabla S_e -i \hbar s^* \nabla s +(e/c) {\bf A}$.

We can now express the quantum electron velocity \cite{Brodin07R}
\begin{equation}
{\bf v}_e = \frac{\hbar}{2 m_e}\frac{(\Psi^*\nabla\Psi-\Psi\nabla\Psi^*)}{|\Psi|^2}
+ i \frac{\hbar}{m_e} s^* \nabla s -\frac{e}{m_e c}{\bf A},
 \label{velocity}
\end{equation}
the spin density vector
\begin{equation}
 {\bf s}=\frac{\hbar}{2} s^* {\boldsymbol \sigma} s.
\end{equation}
and the spin vector transport equation \cite{Takabayashi52}
\begin{equation}
  \frac{d{\bf s}}{dt}=\frac{e}{m_ec}({\bf s}\times {\bf B})+\frac{1}{m_e n_e}
\bigg[{\bf s}\times\frac{\partial}{\partial x_k}\bigg(n_e\frac{\partial {\bf s}}
{\partial x_k}\bigg)\bigg],
\end{equation}
where we have used the summation convention for repeated indices, and have denoted
$d/dt\equiv (\partial/\partial t) +{\bf v}_e\cdot \nabla$ is the total derivative.
The electron momentum and electromagnetic fields are coupled via the Maxwell equations.

Ignoring the electromagnetic fields and the particle spin, one can obtain, after
linearizing (33) and (34), the frequency of electron oscillations

\begin{equation}
\omega_g = \frac{\hbar k^2}{2m_e},
\end{equation}
where $k$ is the wave number.

Finally, we consider interaction of an electron with both background electrons and singly
charged positive ions. The electron dynamics is governed by \cite{Dvornikov09}

\begin{equation}
i\hbar \frac{\partial \psi}{\partial t} + \frac{\hbar^2}{2m_e}\nabla^2 \psi
-U(|\psi|^2) =0,
\end{equation}
where

\begin{equation}
U(|\psi|^2) = e^2 \int d^3 {\bf r}^{\prime}\frac{1}{|{\bf r}-{\bf r}^{\prime}|}
\left(|\psi({\bf r}^{\prime},t))|^2-n_i ({\bf r}^{\prime},t)\right),
\end{equation}
is the potential responsible for the interaction of an electron with background matter which
includes electrons and positively charged ions with the number density $n_i({\bf r}^{\prime},t)$.
The wave function is normalized on the number density of electrons, viz. $n_e({\bf r}, t)
=|\psi({\bf r}, t)|^2$. We note that Eq. (39) accounts only for the Coulomb interactions between electrons
and ions, and completely ignores the quantum statistical pressure, the self-consistent
ambipolar field arising from the charge separation, and the electron spin-1/2 effect.

Assume that $\psi ({\bf r}, t) =\psi_0 + \psi_1 ({\bf r}, t)$, where $|\psi_0|^2= n_0$ represent
the unperturbed electron number density and the perturbation wave function $\psi_1 ({\bf r}, t)$
for spherically symmetric oscillations has the form $(A_k/r) {\rm sin (kr)} \exp(-i \omega t)$,
with $A_k$ being the normalization constant.  Thus, the dispersion relation
deduced from (39) reads \cite{Dvornikov09}
\begin{equation}
k^2 =\frac{\omega m_e}{\hbar}\left[1 \pm \left(1-4\frac{\omega_{pe}^2}{\omega^2}\right)^{1/2}\right].
\end{equation}

The interaction between two electrons participating in spherically symmetric
electron oscillations has been considered in Ref. \cite{Dvornikov09}. The latter
predicts that there would be an effective attraction between electrons
mediated by low energy [given by the minus sign in Eq. (41)] spherically symmetric
oscillations of electrons in a quantum plasma. We can thus have attracting degenerate
electrons in dense plasmas. The underlying physics of electron attraction here seems to
be similar to that of the Cooper pairing of electrons in superconductors in which electrons
close to the Fermi level attract each other due to their interactions with crystal lattice
vibrations (phonon oscillations). The pairs of electrons act more like bosons which can
condensate into the same energy level, contrary to single electrons which are fermions and
must obey the Pauli exclusion principle.

\section{Nonlinear equations for unmagnetized quantum plasmas}

In the preceding section, we have seen that the quantum Madelung fluid description predicts
a diffraction pattern of a single electron or positron. However, collective interactions
between an ensemble of degenerate electrons (Fermions) in dense  plasmas are responsible for
new linear and nonlinear waves and structures.

The quantum $N$-body problem is governed by the Schr\"odinger equation for the $N$-particle wave
function $\psi(q_1,q_2,\ldots, q_N, t)$, where $q_j=({\bf r}_j,s_j)$ is the coordinate (space, spin)
of particle $j$. For identical Fermions, the equilibrium $N$-particle wave function is given by
the Slater determinant \cite{Bransden00}
\begin{equation}
\psi(q_1,\,q_2,\,\ldots,\,q_N, t) = \frac{1}{\sqrt{N!}}
\left| \begin{array}{cccc}
\psi_1(q_1,t)& \psi_2(q_1,t)& \cdots &\psi_N(q_1,t)\\
\psi_1(q_2,t)& \psi_2(q_2,t)& \cdots &\psi_N(q_2,t)\\
\vdots & \vdots& \ddots & \vdots\\
\psi_1(q_N,t)& \psi_2(q_N,t)& \cdots &\psi_N(q_N,t)\\
\end{array}\right|,
\end{equation}
which is anti-symmetric under odd numbers of permutations. Hence, $\psi$ vanishes if two rows are identical,
which is an expression of the Pauli exclusion principle that two identical Fermions cannot occupy the same state.
Example $(N=2)$: $\psi(q_1,q_2,t)=\frac{1}{\sqrt{2}}[\psi_1(q_1,t)\psi_2(q_2,t)-\psi_1(q_2,t)\psi_2(q_1,t)]$
so that $\psi(q_2,q_1,t)=-\psi(q_1,q_2,t)$ and $\psi(q_1,q_1,t)=0$. Due to the Pauli exclusion principle,
all electrons are not permitted to occupy the lowest energy state, and in the ultra-cold limit when all energy
states up to the Fermi energy level are occupied by electrons, there is still a quantum-statistical pressure
determined by the Fermi pressure.

To describe collective electrostatic oscillations in a plasma, the quantum analogue of the Vlasov-Poisson system is the Wigner-Poisson system, given by
\begin{equation}
\frac{\partial f}{\partial t} + {\bf v}\cdot \nabla f= -\frac{iem_e^3}{(2\pi)^3  \hbar^4}
\int\!\!\!\int e^{im_e({\bf v}-\bf{v}')\cdot {\boldsymbol\lambda}/\hbar}
\bigg[\phi\bigg({\bf x }+\frac{\boldsymbol
\lambda}{2},t\bigg)-\phi\bigg({\bf x}-\frac{\boldsymbol
\lambda}{2},t\bigg)\bigg]f({\bf x},{\bf v}',t)\, d^3\lambda\, d^3v',
\label{eq43}
\end{equation}
and
\begin{equation}
\nabla^2 \phi =4 \pi e\left(\int f d^3v -n_0\right),
\end{equation}
assuming immobile ions. We note (see Appendix A) that the Wigner equation converges to the Vlasov equation
for classical particles (electrons) when $\hbar\rightarrow 0$
\begin{equation}
  \frac{\partial f}{\partial t} + {\bf v}\cdot \nabla f
=- \frac{e}{m_e}\nabla\phi\cdot\frac{\partial f}{\partial {\bf v}}.
\end{equation}

We now take the moments of the Wigner equation (\ref{eq43}) and obtain [up to $O(\hbar^2)$] the non-relativistic
quantum-electron fluid (or the quantum Madelung fluid) equations \cite{Manfredi01,Manfredi05} composed of the
electron continuity equation
\begin{equation}
  \frac{\partial n_e}{\partial t} + \nabla \cdot (n_e {\bf u}_e) =0
\end{equation}
the electron momentum equation including the quantum statistical pressure and the quantum force
\begin{equation}
 m_e\left(\frac{\partial }{\partial t} + {\bf u}_e\cdot \nabla\right) {\bf u}_e
=e\nabla \phi -\frac{1}{n_e} \nabla P_e+{\bf F}_Q,
\end{equation}
where $\phi$ is determined from the Poisson equation
\begin{equation}
\nabla^2\phi=4\pi e(n_e-n_0).
\end{equation}

For the degenerate Fermi-Dirac distributed plasma, one has (up to constants of order unity) the
quantum statistical pressure for the electrons
\begin{equation}
 P_e = \frac{m_e V_{Fe}^2n_0}{3}\left(\frac{n_e}{n_0}\right)^{(D+2)/D},
\end{equation}
where $D$ is the number of degrees of freedom in the system.  It should be noted that
Eliasson and Shukla \cite{Eliasson08a} and Tsintsadze and Tsintsadze \cite{Tsintsadze09}
have obtained different expressions for $P_e$ in the non-zero limit of the
electron Fermi temperature.

The quantum force \cite{Gardner96} due to electron tunneling through the Bohm potential is
\begin{equation}
{\bf F}_Q=\frac{\hbar^2}{2m_e}\nabla
\left(\frac{\nabla^2\sqrt{n_e}}{\sqrt{n_e}}\right) \equiv-\nabla \phi_B,
\end{equation}
where $\phi_B$ represents the Bohm potential. We note that the $\alpha$-term [cf. (14)] does not
appear in (50) due to consideration of the term up to $O(\hbar^2)$ in the expansion parameter, as
it also happens when the mean-field approximation \cite{Gardner96} is used in deducing ${\bf F}_Q$.

\subsection{Nonlinear Schr\"odinger-Poisson equations}

By introducing the wave function
\begin{equation}
\psi ({\bf r}, t) =\sqrt{n_e ({\bf r}, t)}\exp(i \varphi_e({\bf r}, t)/\hbar),
\end{equation}
where $S$ is defined according to $m_e {\bf u}_e =\nabla \varphi_e$ and $n_e= |\psi|^2$, it can
be shown that the QHD equations [e.g. Eqs. (46)--(48)] are equivalent to the generalized NLS-Poisson
system  \cite{Manfredi01,Manfredi05}
\begin{equation}
i \hbar \frac{\partial \psi}{\partial t} + \frac{\hbar^2}{2m_e}\nabla^2 \psi
+ e \phi \psi - \frac{m_e V_{Fe}^2}{2n_0^2}|\psi|^{4/D} \psi =0,
  \label{NLS}
\end{equation}
and
\begin{equation}
\nabla^2 \phi = 4\pi e(|\psi^2| - n_0).
\label{Poisson}
\end{equation}
The derivation of (52) required the electron plasma flow velocity to be curl free everywhere in a singly-connected region, except at points where the electron number density vanishes. This is obviously not valid in general.
Similar to quantum-fluid treatment, one should include the generalized electron vorticity
${\boldsymbol \Omega}_e=\nabla\times(|\psi|^2 {\bf u}_e)/|\psi|^2$,
which is non-vanishing [see the discussions below Eq. (25)]. Equation (\ref{NLS}) captures the two
main properties of a quantum plasma, namely
the quantum statistical pressure (fully nonlinear) and quantum dispersion effects, and is coupled
self-consistently to the electrostatic potential via the Poisson equation (\ref{Poisson}). We thus
have a nonlocal nonlinear interaction between the electron density and the electrostatic potential.
Furthermore, we note that one-dimensional version of Eq. (\ref{NLS}) without the $\phi$-term has also
been used to describe the behavior of a Bose-Einstein condensate \cite{r12} in the absence of the
confining potential.

Linearization of the NLS-Poisson Equations yields the frequency of the
EPOs \cite{Klimontovich52,Bohm52,Pines1,Pines2}
\begin{equation}
\omega_k=\left(\omega_{pe}^2+k^2V_{Fe}^2+\frac{\hbar^2k^4}{4m_e^2}\right)^{1/2}.
\end{equation}
One can identify two distinct dispersive effects from (54): One long wavelength regime,
$ V_{Fe}\gg \hbar k/2 m_e $, and one short wavelength regime, $ V_{Fe}\lesssim \hbar k/2 m_e,$
separated by a critical wavenumber $k_{crit}=2\pi/\lambda_{crit}=\pi\hbar/m_e V_{Fe} \sim n_e^{-1/3}$.

\subsection{Inclusion of the ion dynamics}

The dynamics of low-phase speed (in comparison with the electron Fermi speed) nonlinear
electrostatic ion oscillations in a quantum electron-ion plasma is governed by the inertialess electron
equation of motion
\begin{equation}
0 =e\nabla \phi -\frac{1}{n_e} \nabla P_e + {\bf F}_Q,
\label{electron}
\end{equation}
the ion continuity
\begin{equation}
\frac{\partial n_i}{\partial t} + \nabla \cdot (n_i {\bf u}_i ) =0,
\label{ioncontinuity}
\end{equation}
the ion momentum equation
\begin{equation}
m_i n_i \left(\frac{\partial}{\partial t} + {\bf u}_i \cdot \nabla \right) {\bf u}_i
=-Z_i n_i e \nabla \phi
\label{ionmomentum}
\end{equation}
and the Poisson equation
\begin{equation}
\nabla^2\phi = 4 \pi e(n_e -Z_i n_i),
\end{equation}
where $n_i$ is the ion number density, ${\bf u}_i$ is the ion fluid velocity, and $Z_i$
is the ion charge state.

In the quasi-neutral approximation, viz. $n_i =Z_i n_i =n$, we can combine Eqs. (\ref{electron}) and
(\ref{ionmomentum}) to obtain
\begin{equation}
\left(\frac{\partial}{\partial t} + {\bf u}_i \cdot \nabla \right) {\bf u}_i
=- \frac{C_{Fs}^2 n_0}{n_i} \nabla \left(\frac{Z_i n_i}{n_0}\right)^{(D+2)/D}
+ \frac{\hbar^2}{2m_e m_i} \nabla\left(\frac{\nabla^2 \sqrt{Z_i n_i}}{\sqrt{Z_i n_i}}\right).
\label{ionmomentum1}
\end{equation}
Equations (\ref{ioncontinuity}) and (\ref{ionmomentum1}) are the desired set for studying
nonlinear ion waves \cite{Haas03,Eliasson08} in a dense quantum plasma.

\section{Localized electrostatic excitations}

We are now in a position to discuss nonlinear properties and the dynamics of localized electrostatic
excitations in a dense quantum plasma based on the nonlinear equations we have developed in
the preceding section.

\subsection{Dark solitons and vortices associated with EPOs
\label{sec:vortices}}

In this subsection, we shall discuss the formation of one-dimensional quantized dark solitons
and two-dimensional (2D) quantized vortices associated with the EPOs in a dense quantum plasma
at nanoscales (of the order of $V_{Fe}/\omega_{pe}$). We note that the dynamics of dark
solitons in the 2D nonlinear Schr\"odinger equation in a defocusing medium has been studied
by Ivonin \cite{Ivonin97} and Ivonin et al. \cite{Ivonin99}.  However, for studying the formation
and dynamics of electrostatic nanostructures in a dense quantum plasma, we shall use the
nonlinear Schr\"odinger-Poisson equations \cite{Shukla06}

\begin{equation}
i \frac{\partial \Psi}{\partial t} + A  \nabla^2 \Psi + \varphi \Psi
- |\Psi|^{4/D} \Psi =0,
\label{Eq1}
\end{equation}
and
\begin{equation}
\nabla^2\varphi = |\Psi|^2 -1,
\label{Eq2}
\end{equation}
where the normalized wave function $\Psi$ is $\psi/\sqrt{n_0}$, the normalized $\varphi$
is $e\phi/k_B T_{Fe}$, and the time and space variables are in units of $\hbar/k_B T_{Fe}$ and
$V_{Fe}/\omega_{pe}$, respectively. We have denoted $A= 4\pi m_e e^2/\hbar^2 n_0^{1/3}$
(see Ref. \cite{Shukla06}).

The system of equations(\ref{Eq1}) and (\ref{Eq2}) is supplemented by the Maxwell equation
\begin{equation}
\frac{\partial {\bf E}}{\partial t}=
iA\left(\Psi\nabla\Psi^*-\Psi^*\nabla\Psi\right),
\label{Eq3}
\end{equation}
where the electric field is ${\bf E}=-\nabla \phi$.

The system (\ref{Eq1})--(\ref{Eq3}) has the following conserved integrals:
the number of electrons
\begin{equation}
  N=\int |\Psi|\, d^3 x,
\end{equation}
the electron momentum
\begin{equation}
{\bf P}=-i\int \Psi^*\nabla\Psi\, d^3 x,
\end{equation}
the electron angular momentum
\begin{equation}
  {\bf L}=-i\int \Psi^*{\bf r}\times\nabla\Psi\, d^3 x,
\end{equation}
and the total energy
\begin{equation}
{\cal E}=\int [-\Psi^*A\nabla^2\Psi + |\nabla\varphi|^2/2 + |\Psi|^{2+4/D}D/(2+D)] \,d^3 x.
\end{equation}
The conserved quantities are required to check the accuracy of the numerical integration of
(60) and (61).

For quasi-stationary one-dimensional structures moving with a constant speed $v_0$, one can
find localized, solitary wave solutions by introducing the ansatz $\Psi=W(\xi)\exp(iKx-i\Omega t)$,
where $W$ is a complex-valued function of the argument $\xi=x-v_0t$, and $K$ and $\Omega$ are a
constant wavenumber and the frequency shift, respectively. By the choice $K=v_0/2A$, the coupled
system of equations can be written as
\begin{equation}
 \frac{d^2 W}{d\xi^2}+\lambda W+\frac{\varphi W}{A}-\frac{|W|^4
W}{A}=0,
\label{Eq4}
\end{equation}
and
\begin{equation}
\frac{d^2\varphi}{d\xi^2}=|W|^2-1,
\label{Eq5}
\end{equation}
where $\lambda=\Omega/A-v_0^2/4A^2$ is an eigenvalue of the system.
From the boundary conditions $|W|=1$ and $\varphi=0$ at $|\xi|=\infty$,
we determine $\lambda=1/A$ and $\Omega=1+v_0^2/4A$. The system of
Eqs. (\ref{Eq4}) and (\ref{Eq5}) admits a first integral in the form
\begin{equation}
{\cal  H} =A\left|\frac{dW}{d\xi}\right|^2
  -\frac{1}{2}\left(\frac{d\varphi}{d\xi}\right)^2
  +|W|^2
  -\frac{|W|^6}{3}+\varphi |W|^2-\varphi-\frac{2}{3}=0,
  \label{Eq6}
\end{equation}
where the boundary conditions $|W|=1$ and $\varphi=0$
at $|\xi|=\infty$ have been employed.

\begin{figure}[htb]
\centering
\includegraphics[width=8cm]{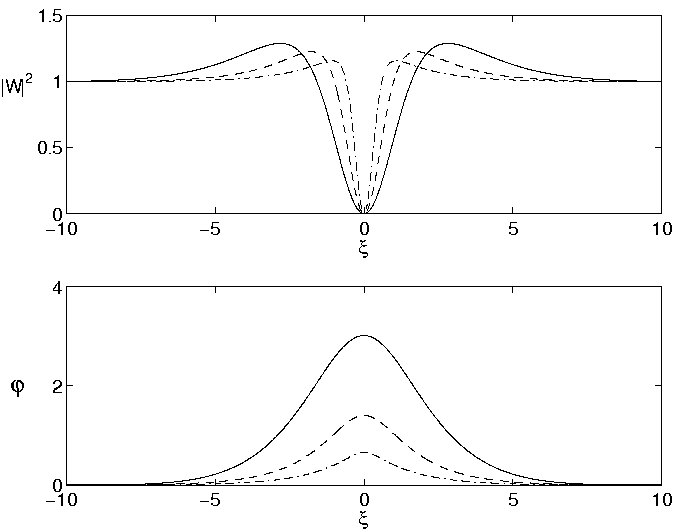}
\caption{The electron density $|W|^2$ (the upper panel) and
electrostatic potential $\varphi$ (the lower panel) associated with a
dark soliton supported by the system of equations (\ref{Eq4}) and (68), for $A=5$ (solid
lines), $A=1$ (dashed lines), and $A=0.2$ (dash-dotted line). After Ref.~\cite{Shukla06}.}
\label{Fig1}
\end{figure}

\begin{figure}[htb]
\centering
\includegraphics[width=8cm]{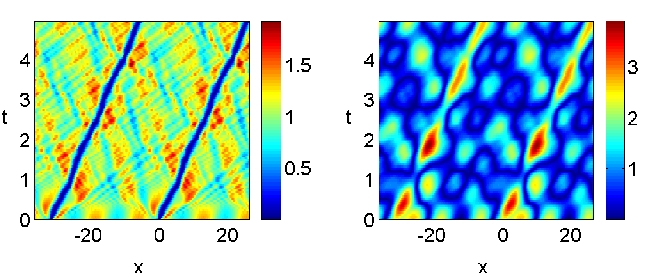}
\caption{The time-development of the electron density $|\Psi|^2$
(left-hand panel) and electrostatic potential $\varphi$ (the right-hand
panel), obtained from a simulation of the system of equations (\ref{Eq1}) and (\ref{Eq2}).
The initial condition is $\Psi=0.18+\tanh[20\sin(x/10)]\exp(i K x)$,
with $K=v_0/2A$, $A=5$ and $v_0=5$.  After Ref.~\cite{Shukla06}.}
\label{Fig2}
\end{figure}

Figure \ref{Fig1} shows profiles of $|W|^2$ and
$\varphi$ obtained numerically from (\ref{Eq4}) and (\ref{Eq5})
for a few values of $A$, where $W$ was set to $-1$ on the
left boundary and to $+1$ on the right boundary, i.e. the phase
shift is 180 degrees between the two boundaries. The solutions are in the form
of dark solitons, with a localized depletion
of the electron density $N_e=|W|^2$, associated with a localized positive
potential. Larger values of the parameter quantum coupling parameter
$A$ give rise to larger-amplitude and wider dark solitons. The solitons
localized ```shoulders'' on both sides of the density depletion.

Numerical solutions of the time-dependent system of Eqs. (\ref{Eq1})
and (\ref{Eq2}) is displayed in Fig. \ref{Fig2}, with initial conditions
close (but not equal) to the ones in Fig. \ref{Fig1}.
Two very clear and long-lived dark solitons are visible, associated with a positive
potential of $\varphi\approx 3$, in agreement with the
quasi-stationary solution of Fig. \ref{Fig1} for $A=5$. In addition
there are oscillations and wave turbulence in the
time-dependent solution presented in Fig. \ref{Fig2}.
Hence, the dark solitons seem to be robust structures that can withstand
perturbations and turbulence during a considerable time.

\begin{figure}[htb]
\centering
\includegraphics[width=8cm]{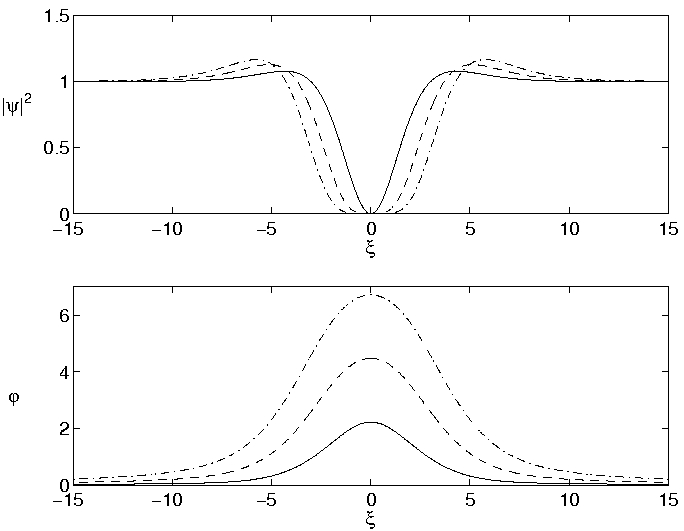}
\caption{The electron density $|\Psi|^2$ (upper panel) and
electrostatic potential $\varphi$ (lower panel) associated with a
two-dimensional vortex supported by the system (\ref{Eq7}) and (\ref{Eq8}), for the
charge states $s=1$ (solid lines), $s=2$ (dashed lines) and $s=3$
(dash-dotted lines). We used $A=5$ in all cases. After Ref.~\cite{Shukla06}.}
\label{Fig3}
\end{figure}

\begin{figure}[htb]
\centering
\includegraphics[width=8cm]{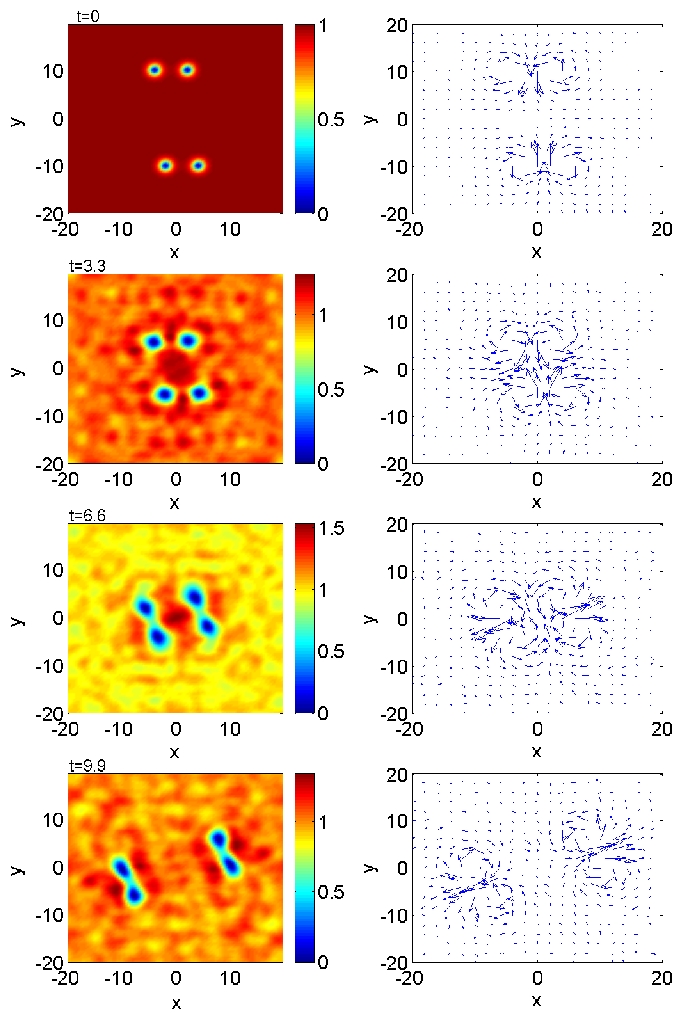}
\caption{The electron density $|\Psi|^2$ (left panel) and an arrow
plot of the electron current
$i\left(\Psi\nabla\Psi^*-\Psi^*\nabla\Psi\right)$ (right panel)
associated with singly charged ($|s|=1$) two-dimensional vortices,
obtained from a simulation of the time-dependent system of equations
(\ref{Eq1}) and (\ref{Eq2}), at times $t=0$, $t=3.3$, $t=6.6$ and $t=9.9$ (upper to
lower panels). We used $A=5$. The singly charged vortices form pairs
and keep their identities. After Ref.~\cite{Shukla06}.}
\label{Fig4}
\end{figure}

\begin{figure}[htb]
\centering
\includegraphics[width=8cm]{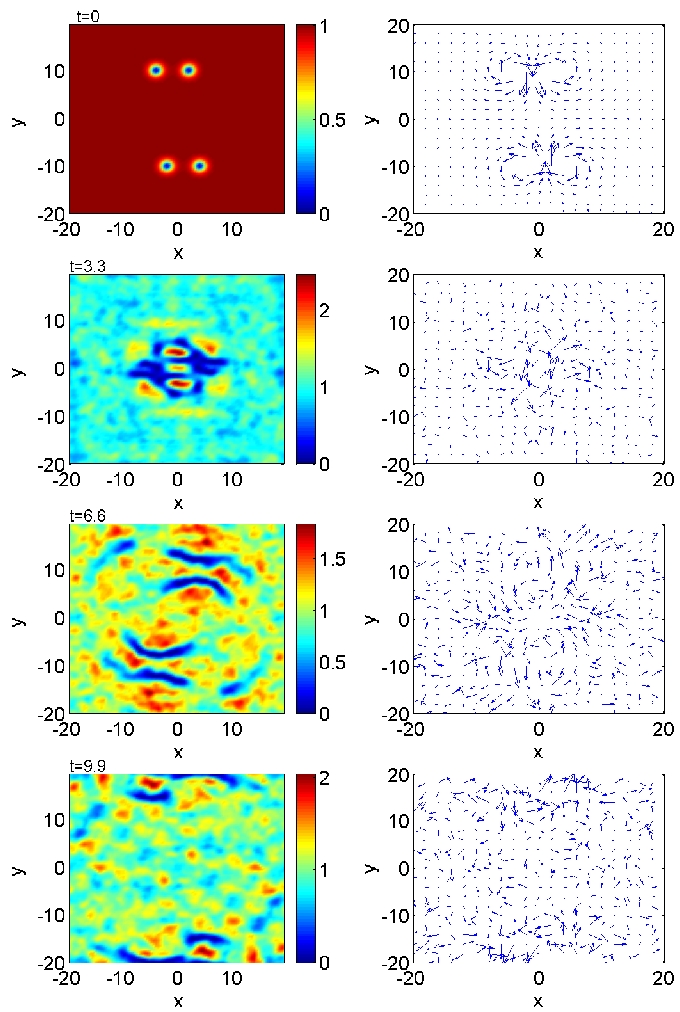}
\caption{The electron density $|\Psi|^2$ (left panel) and an arrow
plot of the electron current
$i\left(\Psi\nabla\Psi^*-\Psi^*\nabla\Psi\right)$ (right panel)
associated with double charged ($|s|=2$) two-dimensional vortices,
obtained from a simulation of the time-dependent system of Eqs.
(\ref{Eq1}) and (\ref{Eq2}), at times $t=0$, $t=3.3$, $t=6.6$ and $t=9.9$ (upper to
lower panels). We used $A=5$. The doubly charged vortices dissolve
into nonlinear structures and wave turbulence. After Ref.~\cite{Shukla06}.}
\label{Fig5}
\end{figure}

For the two-dimensional ($D=2$) system, it is possible to find vortex structures
of the form $\Psi=\psi(r)\exp(is\theta-i\Omega t)$, where $r$ and $\theta$ are
the polar coordinates defined via $x=r\cos(\theta)$ and
$y=r\sin(\theta)$, $\Omega$ is a constant frequency shift, and $s=0,
\, \pm 1,\, \pm 2,\ldots$ for different excited states (charge
states). The index $s$ is also known as the circulation number \cite{Ghosh82}.
With this ansatz, Eqs. (\ref{Eq1}) and (\ref{Eq2}) can be written in the form
\begin{eqnarray}
&&\!\!\!\!\!\!\!\!\!\!\!\!\!\!\!\!\!\!\!\!\!\!\!\!\!\!\!
\left[\Omega+A\left(\frac{d^2}{dr^2}+\frac{1}{r}\frac{d}{dr}-\frac{s^2}{r^2}\right)
+\varphi-|\psi|^2\right]\psi=0,
\label{Eq7}
\end{eqnarray}
and
\begin{equation}
\left(\frac{d^2}{dr^2}+\frac{1}{r}\frac{d}{dr}\right)\varphi=|\psi|^2-1,
\label{Eq8}
\end{equation}
respectively, where the boundary conditions $\psi=1$ and
$\varphi=d\psi/dr=0$ at $r=\infty$ determine the constant frequency $\Omega=1$.
Different signs of charge state $s$ describe different rotation directions of the vortex.
For $s\neq 0$, one must have $\psi=0$ at $r=0$, and from symmetry
considerations one has $d\varphi/dr=0$ at $r=0$. Figure \ref{Fig3} shows
numerical solutions of Eqs. (\ref{Eq7}) and (\ref{Eq8}) for different
$s$ and for $A=5$. Here the vortex is characterized by
a complete depletion of the electron density at the core of the
vortex, and is associated with a positive electrostatic potential.

Figure \ref{Fig4} shows time-dependent solutions of Eqs. (\ref{Eq1}) and (\ref{Eq2}) in
two space dimensions for singly charged ($s=\pm 1$) vortices, where, in the initial
condition, four vortex-like structures were placed at some distance from each other.
The initial conditions were such that the vortices are organized in two vortex
pairs, with $s_1=+1$, $s_2=-1$, $s_3=-1$, and $s_4=+1$, seen
in the upper panels of Fig. \ref{Fig4}. The vortices in the
pairs have opposite polarity on the electron fluid rotation,
as seen in the in the upper right panel of Fig. \ref{Fig4}. Interestingly,
the ``partners'' in the vortex pairs attract each other and
propagate together with a constant velocity, and in the collision and interaction
of the vortex pairs (see the second and third pairs of panels in Fig. \ref{Fig4}), the vortices
keep their identities and change partners, resulting into two new vortex pairs which propagate
obliquely to the original propagation direction. On the other hand,
as shown in Fig. \ref{Fig5}, vortices that are multiply charged ($|s_j|>1$) are unstable.
Here the system of Eqs. (\ref{Eq1}) and (\ref{Eq2}) was again solved numerically with
the same initial condition as the one in Fig. \ref{Fig4}, but with doubly charged vortices
$s_1=+2$, $s_2=-2$, $s_3=-2$, and $s_4=+2$. The second row
of panels in Fig. \ref{Fig5} reveals that the vortex pairs keep their
identities for some time, while a quasi one-dimensional density
cavity is formed between the two vortex pairs. At a later stage, the
four vortices dissolve into complicated nonlinear structures and
wave turbulence. Hence, the nonlinear dynamics is very different
between singly and multiply charged solitons, where only singly
charged vortices are long-lived and keep their identities. This is
in line with previous results on the nonlinear Schr\"odinger
equation, where it was noted that vortices with higher charge states
are unstable \cite{Ivonin99}.

\subsection{Localized ion wave excitations in  quantum plasmas}

In his classic paper, Haas et al. \cite{Haas03} developed both small and large
amplitude theories for one-dimensional solitary ion waves in a dense quantum
plasmas. They found that the dynamics of small amplitude solitary waves is
governed by the Kortweg de-Vries (k-dV) equation
\begin{equation}
\frac{\partial U}{\partial \tau} + 2 U \frac{\partial U}{\partial \xi}
+ \frac{1}{2}\left(1-\frac{H}{8}\right) \frac{\partial^3U}{\partial \xi^3} =0,
\label{qiaEq0}
\end{equation}
where $U$ represents the relative (with respect to $n_0$) ion density perturbation,
the time and space variables are in units of
the ion plasma period $\omega_{pi}^{-1}$ and the electron Thomas-Fermi radius
$(k_B T_{Fe}/4\pi n_0 e^2)^{1/2}$, respectively.

The K-dV equation admits both the solitary and the periodic (cnoidal) waves \cite{Whitham99}.
Introducing the wave form $U(\eta=\xi-M_s \tau)$, we can write (72) in the stationary
frame as
\begin{equation}
\frac{1}{2}\left(1-\frac{H}{8}\right) \frac{\partial^2U}{\partial \eta^2}- M_s U + U^2 + C=0,
\label{qiaEq00}
\end{equation}
where $M_s$ represents the Mach number, and $C$ is a constant of integration. In the special
case when $U$ and its derivatives tend to zero at $\pm \infty$, $C=0$. Multiplying (73) by
$\partial U/\partial \eta$ one can integrate once the resultant equation, and express it in
the form of an energy integral \cite{Sagdeev66,Sagdeev79,Shukla78a}. The resulting solitary wave
solution of (73) is
\begin{equation}
U = U_m {\rm sech}^2 (\eta/\eta_0),
\label{qiaEq000}
\end{equation}
where $U_m =(3M_s/2)$ and $\eta_0=(2/M_s)^{1/2}(1-H/8)^{1/2}$ are the maximum amplitude
and the width of the soliton. We see that compressive solitary wave solutions are possible
if $0 < H < 8$.

There also exist possibility of one-dimensional large amplitude localized ion wave excitations.
To demonstrate this, we  assume that the quantum force acting on the electrons dominates
over the quantum statistical pressure, viz. $k_BT_{Fe}n_e \ll (\hbar^2/4m_e)
\partial^2 n_{e}/\partial x^2$. Hence, the electron density is obtained from \cite{Eliasson08}

\begin{equation}
  e \phi +\frac{\hbar^2}{2 m_e \sqrt{n_e}} \frac{\partial^2 \sqrt{n_e}}{\partial x^2} =0.
  \label{qiawEq1}
\end{equation}
The electrons are coupled with ions through the space charge electric field ($-\nabla \phi$).

The dynamics of singly charged ions is governed by the ion continuity
\begin{equation}
  \frac{\partial n_i}{\partial t}+ \frac{\partial(n_i u_i)}{\partial x}=0,
  \label{qiawEq2}
\end{equation}
and ion momentum equation
\begin{equation}
  m_i\left(\frac{\partial}{\partial t}+ u_i \frac{\partial}{\partial x}\right) u_i=-
e \frac{\partial \phi}{\partial x},
  \label{qiawEq3}
\end{equation}
where $u_i$ is $x$ component of the ion fluid velocity perturbation.
The system of Eqs. (\ref{qiawEq1})--(\ref{qiawEq3}) is closed by
the Poisson equation

\begin{equation}
\frac{\partial^2 \phi}{\partial x^2} = 4\pi e(n_e-n_i).
  \label{qiawEq4}
\end{equation}

We now look for stationary nonlinear ion wave structures moving with a constant speed $u_0$.
Hence, all unknowns depend only on the variable $\xi=x-u_0 t$.
Defining $\sqrt{n_e}\equiv\psi$, Eq. (\ref{qiawEq1}) takes the form
\begin{equation}
  \frac{\hbar^2}{2 m_e}\frac{\partial^2\psi}{\partial \xi^2}+e\phi\psi=0.
  \label{qiawEq5}
\end{equation}

Equations (\ref{qiawEq2}) and (\ref{qiawEq3}) can be integrated once with the boundary conditions
$n_i=n_0$ and $u_i=0$ at $\xi=|\infty|$, and the results can be combined to have
\begin{equation}
  n_i=\frac{n_0u_0}{\sqrt{u_0^2-2e\phi/m_i}}.
  \label{qiawEq6}
\end{equation}
Inserting (\ref{qiawEq6}) into (\ref{qiawEq4}) we obtain
\begin{equation}
  \frac{\partial^2\phi}{\partial \xi^2}=4\pi e\left(
    \psi^2-\frac{n_0u_0}{\sqrt{u_0^2-2e\phi/m_i}} \right).
  \label{qiawEq7}
\end{equation}
Equations (\ref{qiawEq5}) and (\ref{qiawEq7}) are the desired equations for studying the
nonlinear ion waves in dense quantum plasmas.

It is convenient to introduce dimensionless quantities [see below Eq. (\ref{qiawEq9})] into
Eqs. (\ref{qiawEq5}) and (\ref{qiawEq7}), and rewrite them as
\begin{equation}
  \frac{\partial^2\Psi}{\partial X^2}+\frac{\Phi\Psi}{2}=0,
  \label{qiawEq8}
\end{equation}
and
\begin{equation}
  \frac{\partial^2\Phi}{\partial X^2}-
    \Psi^2+\frac{M}{\sqrt{M-2\Phi}}=0,
    \label{qiawEq9}
\end{equation}
where we have normalized the space variable as $X=k_q \xi$, the
electron wave function as $\Psi=\sqrt{n_0}\psi$, and the potential
as $\Phi=e\phi/m_i c_q^2$. Here $c_q=\omega_{pi}/k_q$ is the quantum
ion wave speed and $k_q=(2m_e\omega_{pe}/\hbar)^{1/2}$ is the
quantum wavenumber. The quantum "Mach number" is defined as $M=u_0/c_q$.

We note that the coupled Eqs. (\ref{qiawEq8}) and (\ref{qiawEq9}) admit a conserved quantity
\begin{equation}
  {\cal H}=-2\left(\frac{\partial\Psi}{\partial X}\right)^2
  +\frac{1}{2}\left(\frac{\partial\Phi}{\partial X}\right)^2
  -\Phi\Psi^2-M(\sqrt{M^2-2\Phi}-M)=0,
\label{qiawEq10}
\end{equation}
where we have used the boundary conditions $\Phi=\partial\Phi/\partial X=\partial\Psi/\partial X=0$,
and $\Psi=1$ at $|X|=\infty$. For a symmetric solitary ion wave structure, we can assume that
$\Phi=\Phi_{\rm max}$ and $\Psi=\Psi_{\rm max}$, as well as
${\partial\Psi}/{\partial X}={\partial\Phi}/{\partial X}=0$ at $X=0$.
Hence, at $X=0$ Eq. (\ref{qiawEq10}) yields
\begin{equation}
\Phi_{\rm max}\Psi_{\rm max}^2 + M(\sqrt{M^2-2\Phi_{\rm max}}-M)=0.
\end{equation}
In the wave-breaking limit, where $M=(2\Phi_{\rm max})^{1/2}$,
we find that $-\Phi_{\rm max}\Psi_{\rm max}^2+ 2\Phi_{\rm max}=0$, or
$\Psi_{\rm max}=2$. Accordingly, the electron density will locally rise to twice the
background density at wave breaking.

\begin{figure}[htb]
\centering
\includegraphics[width=8.5cm]{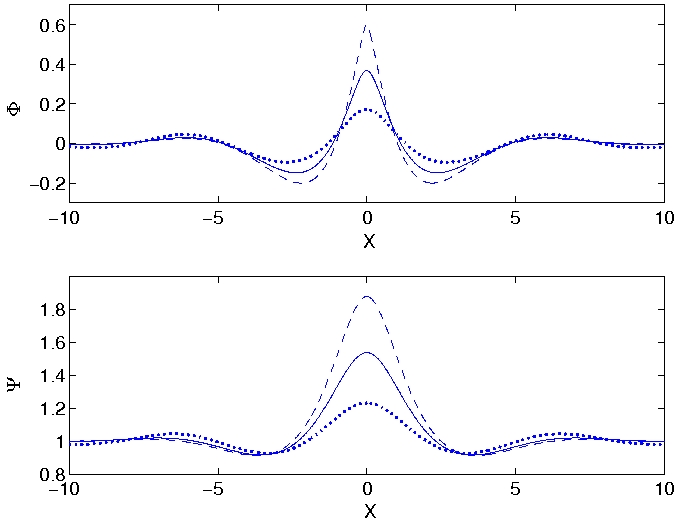}
\caption{The profiles of the potential $\Phi$ (top panel) and the electron
density $\Psi^2$ (bottom panel) as a function of $X$,
for different values of the Mach number: $M=1.1$ (dashed curves), $M=0.9$ (solid curve), and
$M=0.75$ (dotted curve). After Ref. \cite{Eliasson08}.}
\label{qiawFig1}
\end{figure}

In Fig. \ref{qiawFig1}, numerical solutions of Eqs. (\ref{qiawEq8}) and (\ref{qiawEq9}) are displayed,
showing profiles of the electrostatic potential and electron number densities for different values
of $M$.  We see that both the electrostatic potential and electron density have localized and strongly
peaked maxima and an oscillatory tail. The latter is in sharp contrast to the classical (non-quantum) case,
where the ion acoustic solitary waves have a monotonic profile, which in the small amplitude limit,
where the system is governed by the Korteweg-de Vries equation, assumes a secant hyperbolicus shape.
We observe from Fig. \ref{qiawFig1} that $M=1.1$ is close to the wave breaking limit above which there
do not exist solitary wave solutions. Our numerical investigation also suggests that there is a lower
limit of $M$ (slightly lower than 0.75), below which the solitary wave solution vanishes.

\section{Quantum fluid turbulence\label{sec:turbulence}}
The statistical properties of turbulence and its associated electron transport
at nanoscales in quantum plasmas has been investigated in both 2D and 3D by means
of the coupled NLS and Poisson equations \cite{Shaikh07,Shaikh08}. It has been found
that the nonlinear coupling between the EPOs of different scale sizes gives rise to
small-scale electron density structures, while the electrostatic potential cascades
towards large-scales. The total energy associated with the quantum electron plasma
turbulence, nonetheless, processes a characteristic, {\it non-}Kolmogorov-like spectrum.
The electron diffusion caused by the electron fluid turbulence is consistent with the
dynamical evolution of turbulent mode structures.

To investigate the quantum electron fluid turbulence in 3D, we use the nonlinear
Schr\"odinger-Poisson equations \cite{Manfredi01,Shukla06,Shaikh08}
\begin{equation}
i \sqrt{2H} \frac{\partial \Psi}{\partial t}+ H \nabla^2\Psi
+ \varphi \Psi - |\Psi|^{4/3}\Psi = 0,
\label{Eq1D}
\end{equation}
and
\begin{equation}
\nabla^2\varphi = |\Psi|^2-1,
\label{Eq2D}
\end{equation}
which govern the dynamics of nonlinearly interacting EPOs of different wavelengths.
In Eqs. (\ref{Eq1D}) and (\ref{Eq2D}) the wave function is normalized by $\sqrt{n_0}$, the electrostatic
potential by $k_B T_{Fe}/e$, the time $t$ by the electron plasma period $\omega_{pe}^{-1}$,
and the space ${\bf r}$ by the Thomas-Fermi Debye radius $V_{Fe}/\omega_{pe}$. We have introduced
the notation $\sqrt{H} = \hbar \omega_{pe}/\sqrt{2} k_B T_{Fe}$.

The nonlinear mode coupling interaction studies are performed to investigate the
multi-scale evolution of a decaying 3D electron fluid turbulence, which is described by
Eqs. (\ref{Eq1D}) and (\ref{Eq2D}). All the fluctuations are initialized isotropically
(no mean fields are assumed) with random phases and amplitudes in Fourier space,
and evolved further by the integration of Eqs. (\ref{Eq1D}) and (\ref{Eq2D}), using a fully
de-aliased pseudospectral numerical scheme \cite{Gottlieb77} based on  the Fourier spectral
methods.  The spatial discretization in our 3D simulations uses a discrete Fourier
representation of turbulent fluctuations.  The numerical algorithm
employed here conserves energy in terms of the dynamical fluid
variables and not due to a separate energy equation written in a
conservative form.  The evolution variables use periodic boundary
conditions. The initial isotropic turbulent spectrum was chosen close
to $k^{-2}$, with random phases in all three directions. The choice of
such (or even a flatter than $-2$) spectrum treats the turbulent
fluctuations on an equal footing and avoids any influence on the
dynamical evolution that may be due to the initial spectral
non-symmetry. The equations are advanced in time using a second-order
predictor-corrector scheme. The code is made stable by a proper
de-aliasing of spurious Fourier modes, and by choosing a relatively small
time step in the simulations. Our code is massively parallelized using
Message Passing Interface (MPI) libraries to facilitate higher
resolution in a 3D computational box, with a resolution of $128^3$ grid points.

\begin{figure}[htb]
\centering
\includegraphics[width=8cm]{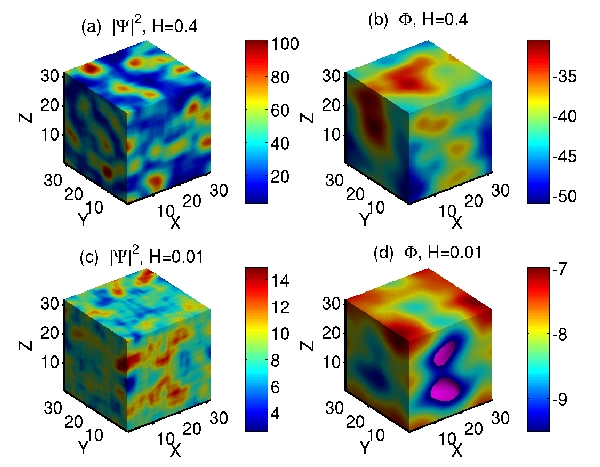}
\caption{Small scale fluctuations in the electron density
resulted from a steady turbulence simulations of our 3D electron
plasma, for $H=0.4$ (top panels) and $H=0.01$ (bottom panels).
Forward cascades are responsible for the generation of
small-scale fluctuations seen in panels (a) and (c).
Large scale structures are present in the
electrostatic potential, seen in panels (b) and (d), essentially
resulting from an inverse
cascade. After Ref.~\cite{Shaikh08}.}
\label{Fig1D}
\end{figure}

\begin{figure}[htb]
\centering
\includegraphics[width=8cm]{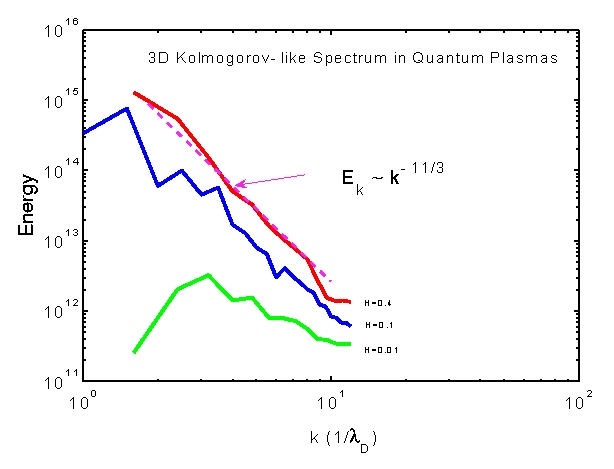}
\caption{
Power spectrum of 3D EPOs in the forward cascade regime.
A Kolmogorov-like spectrum $\sim k^{-11/3}$ is observed for
$H=0.4$. The spectral index changes as a function of $H$.
The numerical resolution is $128^3$.  After Ref.~\cite{Shaikh08}.}
\label{Fig2D}
\end{figure}

We study the properties of 3D fluid turbulence, composed of nonlinearly
interacting EPOs, for two specific physical systems. These
are the dense plasmas in the next generation laser-based plasma compression (LBPC)
schemes \cite{Malkin07} as well as in superdense astrophysical
objects \cite{Chabrier02,Chabrier06,Lai06} (e.g. white dwarfs).
It is expected that in LBPC schemes, the electron number density may reach $10^{27}$ cm$^{-3}$
and beyond. Hence, we have $\omega_{pe} =1.76 \times 10^{18}$ s$^{-1}$, $T_F = 1.7 \times 10^{-9}$
erg, $\hbar \omega_{pe} = 1.7 \times 10^{-9}$ erg, and $H = 1$. The Fermi Debye length $\lambda_D = 0.1$~\AA.
On the other hand, in the interior of white dwarfs, we typically have $n_0 \sim 10^{30}$ cm$^{-3}$,
yielding $\omega_{pe} =5.64 \times 10^{19}$ s$^{-1}$, $T_F = 1.7 \times 10^{-7}$ erg,
$\hbar \omega_{pe} = 5.64 \times 10^{-8}$ erg, $H  \approx 0.3$, and $\lambda_D = 0.025$~\AA.
The numerical solutions of Eqs. (\ref{Eq1D}) and (\ref{Eq2D}) for $H=1$ and $H=0.025$ (corresponding to
$n_0 =10^{27}$ cm$^{-3}$ and $n_0 =10^{30}$ cm$^{-3}$, respectively) are displayed in
Figs. \ref{Fig1D} and \ref{Fig2D}, respectively, which are the electron number density and
electrostatic (ES) potential distributions in the $(x,y)$-plane.

Figures \ref{Fig1D} and \ref{Fig2D} reveal that the electron density
distribution has a tendency to generate smaller length-scale
structures, while the ES potential cascades towards larger scales. The
co-existence of the small and larger scale structures in turbulence is
a ubiquitous feature of various 3D turbulence systems. For example, in
3D hydrodynamic turbulence, the incompressible fluid admits two
invariants, namely the energy and the mean squared vorticity.  The two
invariants, under the action of an external forcing, cascade
simultaneously in turbulence, thereby leading to a dual cascade
phenomena. In these processes, the energy cascades towards longer
length-scales, while the fluid vorticity transfers spectral power
towards shorter length-scales. Usually, a dual cascade is observed in
a driven turbulence simulation, in which certain modes are excited
externally through random turbulent forces in spectral space. The
randomly excited Fourier modes transfer the spectral energy by
conserving the constants of motion in $k$-space. On the other hand, in
freely decaying turbulence, the energy contained in the large-scale
eddies is transferred to the smaller scales, leading to a
statistically stationary inertial regime associated with the forward
cascades of one of the invariants. Decaying turbulence often leads to
the formation of coherent structures as turbulence relaxes, thus
making the nonlinear interactions rather inefficient when they are
saturated.  The power spectrum exhibits an interesting feature in our
3D electron plasma system, unlike the 3D hydrodynamic turbulence
\cite{Kolmogorov41,Lesieur90,Frisch95,Eyink06}. The spectral slope in the 3D quantum
electron fluid turbulence is close to the Iroshnikov-Kraichnan power law
\cite{Iroshnikov63,Kraichnan65} $k^{-3/2}$, rather than the usual Kolomogrov power
law \cite{Kolmogorov41} $k^{-5/3}$. We further find that this scaling is not
universal and is determined critically by the quantum tunneling
effect. For instance, for a higher value of H=1.0 the spectrum becomes
more flat (see Fig \ref{Fig2D}).  Physically, the flatness (or
deviation from the $k^{-5/3}$), results from the short wavelength part
of the EPOs spectrum which is controlled by the quantum tunneling
effect associated with the Bohm potential. The peak in the energy
spectrum can be attributed to the higher turbulent power residing in
the EPO potential, which eventually leads to the generation of larger
scale structures, as the total energy encompasses both the
electrostatic potential and electron density components.  In our dual
cascade process, there is a delicate competition between the EPO
dispersions caused by the statistical pressure law (giving the $k^2
V_F^2$ term, which dominates at longer scales) and the quantum Bohm
potential (giving the $\hbar^2k^4/4m_e^2$ term, which dominates at
shorter scales with respect to a source).  Furthermore, it is
interesting to note that exponents other than $k^{-5/3}$ have also
been observed in numerical simulations \cite{Larichev91,Scott07} of the
Charney and 3D incompressible Navier-Stokes equations. Recently,
Paoletti et al \cite{Paoletti08} have examined the velocity statistics
of quantum turbulence in superfluid $^4He$, and found that
it significantly differs from classical turbulence due to the topological
interactions of vortices that are different from those in classical fluids.

\begin{figure}[htb]
\centering
\includegraphics[width=8cm]{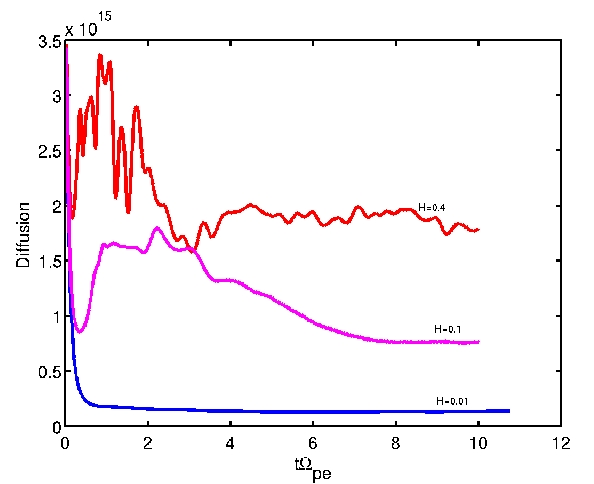}
\caption{Time evolution of an effective electron
diffusion coefficient associated with the large-scale electrostatic
potential and the small-scale electron density, for
$H=0.4$, $H=0.1$ and $H=0.01$. Smaller values of $H$ corresponds to
a small effective diffusion coefficient, which characterizes the
presence of small-scale turbulent eddies that suppress the electron transport.
After Ref.~\cite{Shaikh08}.}
\label{Fig3D}
\end{figure}

We finally estimate the electron diffusion coefficient in the presence
of small and large scale turbulent EPOs in our quantum plasma. An
effective electron diffusion coefficient caused by the momentum
transfer can be calculated from $D_{eff} = \int_0^\infty \langle {\bf
P}({\bf r},t) \cdot {\bf P}({\bf r},t+ t^\prime) \rangle dt^\prime$,
where ${\bf P}$ is electron momentum and the angular bracket denotes
spatial averages and the ensemble averages are normalized to unit
mass. The effective electron diffusion coefficient, $D_{eff}$, essentially
relates the diffusion processes associated with random translational
motions of the electrons in nonlinear plasmonic fields. We compute
$D_{eff}$ in our simulations, to measure the turbulent electron
transport that is associated with the turbulent structures that we
have reported herein. It is observed that the effective electron
diffusion is lower when the field perturbations are Gaussian. On the
other hand, the electron diffusion increases rapidly with the eventual
formation of longer length-scale structures, as shown in Fig. \ref{Fig3D}.
The electron diffusion due to large scale potential distributions in
quantum plasmas dominates substantially, as depicted by the
solid-curve in Fig. \ref{Fig3D}.  Furthermore, in the steady-state, nonlinearly
coupled EPOs form stationary structures, and $D_{eff}$ saturates
eventually. Thus, remarkably an enhanced electron diffusion results
primarily due to the emergence of large-scale potential structures in
our 3D quantum plasma.
\section{Kinetic phase-space structures\label{sec:kinetic}}
In the preceding sections, we have discussed the properties of quantized coherent
structures and 3D quantum electron fluid turbulence based on the coupled Schr\"odinger
and Poisson equations. Thus, it has been assumed that nonlinearly interacting
plasma waves are spontaneously created by some known physical processes
(e.g.  the beam-plasma instability) in  quantum plasmas.

The formation of electrostatic kinetic phase space structures, based on the Vlasov-Poisson
equations, in classical plasmas has been well documented \cite{EliassonShukla06}.
In the following, we shall discuss quasilinear aspects \cite{Haas08b} of the EPOs that
are governed by the Wigner-Poisson system (i.e. a quantum analogue of the Vlasov-Poisson system).
Specifically, we focus on kinetic phase-space nonlinear structures arising from the trapping
of electrons in the finite amplitude wave potential and the self-consistent modification
of the electron distribution function in the presence of nano-kinetic structures.

To study the differences in the nonlinear evolution of the Wigner and Vlasov equations,
we have simulated the well-known bump-on-tail instability \cite{Haas08b}, whereby a
high-velocity beam is used to destabilize a Maxwellian equilibrium. We use the initial
condition $f=(1+\delta)(n_0/\sqrt{2\pi} v_{th}) [0.8 \exp(-v^2/2 v_{th}^2)
+ 0.4 \exp(-2(v-2.5v_{th})^2/v_{th}^2)]$, where $\delta$ represents random
fluctuations of order $10^{-5}$ that help seed the instability
(see Fig. 1). Here $v_{th}=\sqrt{k_B T_e/m_e}$ is the electron
thermal speed. We use periodic boundary conditions with spatial
period $L=40\pi\lambda_{De}$, where $\lambda_{De}= v_{th}/\omega_{p}$
is the Debye length. Three simulations were performed, with different values of the
normalized Planck constant, defined as $H=\hbar\omega_{pe}/mv_{th}^2$: $H=0$ (Vlasov),
$H=1$, and $H=2$.

\begin{figure}[htb]
\includegraphics[width=13.5cm]{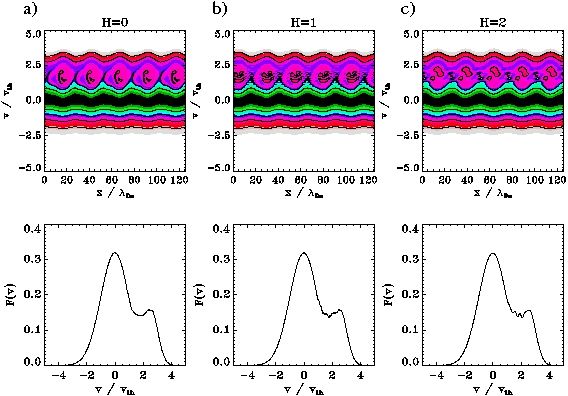}
\caption{(Color online) Simulations of the Wigner-Poisson and Vlasov-Poisson
systems, at time $\omega_{pe} t =200$, for a) $H=0$ (Vlasov),
b) $H=1$ (Wigner), and c) $H=2$ (Wigner). Initially
monochromatic spectrum. Top panels: electron distribution function
$f(x,v)$ in phase space. Bottom panels: spatially averaged
electron distribution function $F(v)$ in velocity space. After Ref.~\cite{Haas08b}.}
\label{Fig2_quasi}
\end{figure}

In order to highlight the transient oscillations in velocity
space, we first perturb the above equilibrium with a monochromatic
wave having $ k \lambda_{De}= 0.25$ (i.e., a wavelength of
$8\pi\lambda_{De}$). Figure \ref{Fig2_quasi} shows the results from simulations
of the Wigner-Poisson and Vlasov-Poisson systems. In both
simulations, due to the bump-on-tail instability, electrostatic
waves develop nonlinearly and create periodic trapped-particle
islands (electron holes) with the wavenumber $k= 0.25 \lambda_{De}^{-1}$.
The theory predicts the formation of velocity-space oscillations in the Wigner
evolution, which should be absent in the classical (Vlasov) simulations. This
is the case in the results presented in Fig. 2, where the
oscillations are clearly visible.

\begin{figure}[htb]
\centering
\includegraphics[width=9cm]{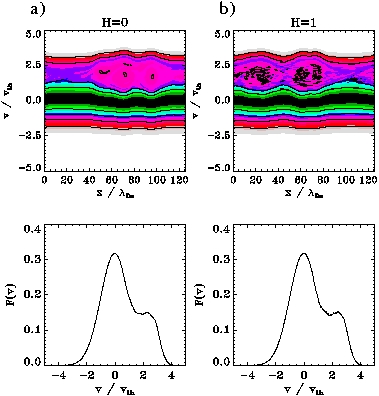}
\caption{(Color online) Simulations of the Wigner-Poisson and Vlasov-Poisson
systems, for $\omega_{pe} t =500$, a) for $H=0$ (Vlasov)
and b) $H=1$ (Wigner). Initially broad wavenumber
spectrum. Top panels: electron distribution function $f(x,v)$ in
phase space. Bottom panels: spatially averaged electron
distribution function $F(v)$ in velocity space. After Ref.~\cite{Haas08b}.}
\label{Fig4_quasi}
\end{figure}

When the initial excitation is broad-band (i.e., wavenumbers $0.05
\le k  \lambda_{De} \le 0.5$ are excited), the electron holes start
merging together at later times due to the sideband instability
\cite{Kruer69,Albrecht07} (see Fig.~\ref{Fig4_quasi}). At this stage,
mode coupling becomes important and quasilinear theory is not capable
of describing these effects. As the system evolves toward larger
spatial wavelength, the evolution becomes progressively more
classical, with the appearance of a plateau in the resonant
region. Nevertheless, at $\omega_{pe} t =500$ the Wigner solution
still displays some oscillatory behavior in velocity space, which
is absent in the Vlasov evolution.

From the experimental viewpoint, recent collective x-ray
scattering observations in warm dense matter \cite{Glenzer07}
revealed a measurable shift in the plasmon frequency due to
quantum effects. In the nonlinear regime (strong excitations),
this effect could lead to trapping of electrons in the wave
potential of the plasmons, and the subsequent formation of the
kind of phase space structures discussed here. Furthermore, we note that
there also exists a theoretical description \cite{Luque04} of quantum
corrected electron holes based on the perturbative treatment of the
Wigner-Poisson equations.

\section{Magnetic fields in quantum plasmas \label{sec:Magnetic}}

There are several mechanisms by which magnetic fields in classical plasmas
can be generated. They include i) non-parallel density and temperature gradients
(the so called Biermann battery \cite{Biermann50}), ii)  the electron temperature
anisotropy (known as the Weibel instability \cite{Weibel59}), iii) counterstreaming
electron beams \cite{Gruzinov01,Schlickeiser03}, and iv)  the ponderomotive forces of
laser beams \cite{Karpman76,Karpman77,Gradov80,Gradov83,Shukla84,Shukla86,Tsintsadze94}.

In the following, we discuss two possibilities for the magnetic field generation
in quantum plasmas.
\subsection{The quantum Weibel instability\label{sec:Weibel}}
We first discuss linear and nonlinear aspects \cite{Levan08,Haas08,HaasLazar08,Haas09} of the Weibel instability
that is driven by equilibrium Fermi-Dirac electron temperature anisotropic distribution function in a nonrelativistic
dense quantum plasma.  It is well known \cite{Pines66} that a dense quantum plasma with an isotropic equilibrium
distribution function does not admit any purely growing linear modes. This can be verified, for instance,
from the expression for the imaginary part of the transverse dielectric function, as derived by
Lindhard \cite{Lindhard54}, for a fully degenerate non-relativistic Fermi plasma. It can be proven
(see Eq. (30) of \cite{Cockayne06}) that the only exception would be for extremely small wavelengths,
so that $k > 2 k_F$, where $k_F$ the characteristic Fermi wave number of the system. However, in
this situation the wave would be super-luminal.  On the other hand, in a classical Vlasov-Maxwell
plasma containing anisotropic electron distribution function, we have a purely growing Weibel
instability \cite{Weibel59}, via which dc magnetic fields are created. The electron temperature
anisotropy may arises due to the heating of the plasma by laser beams \cite{Wei04}, where there
is a signature of the  Weibel instability as well. In the next generation intense laser-solid density
plasma experiments, it is likely that electrons would be degenerate and that electron temperature
anisotropy may develop due to an anisotropic electron heating by intense laser beams via resonant
absorption, similar to the classical laser-plasma-interaction case \cite{Estabrook78}.

Consider linear transverse waves in a dense quantum plasma composed of the  electrons
and immobile ions, with ${\bf k}\cdot{\bf E} = 0$, where ${\bf k}$ is the wave vector
and ${\bf E}$ is the wave electric field.  Following the standard procedure, one then
obtains the general dispersion relation \cite{Klimontovich52,HaasLazar08,Haas09} for the
transverse waves of the Wigner-Maxwell system
\begin{equation}
\omega^2 - \omega_{pe}^2 - k^2 c^2 + \frac{m_e \omega_{pe}^2}{2n_0 \hbar}
\int d{\bf v}\left(\frac{v_{x}^2 + v_{y}^2}{\omega - kv_z}\right)
\left[f_{0}(v_{x},v_{y},v_{z} + \frac{\hbar k}{2m})
- f_{0}(v_{x},v_{y},v_{z} - \frac{\hbar k}{2m_e})\right]= 0,
\label{Eq1weibel}
\end{equation}
where ${\bf v} = (v_{x}, v_{y}, v_{z})$ is the velocity vector, and $f_{0}(v_{x},v_{y},v_z)$
is the equilibrium Wigner function associated to Fermi systems. For spin $1/2$ particles,
the equilibrium pseudo distribution function is in the form of a Fermi-Dirac function.
Here we allow for velocity anisotropy and express
\begin{equation}
f_0 = \frac{\alpha}{\exp\left[\frac{m}{2}\left(\frac{v_{x}^2 + v_{y}^2}{\kappa_{B}T_{\bot}}
+ \frac{v_{z}^2}{\kappa_{B}T_{\parallel}}\right) - \beta\mu\right] + 1} ,
\label{Eq2weibel}
\end{equation}
where $\mu$ is the chemical potential, and the normalization constant is
\begin{equation}
\alpha = - \frac{n_{0}}{{\rm Li}_{3/2}(- e^{\beta\mu})} \Bigl(\frac{m_e\beta}{2\pi}\Bigr)^{3/2}
= 2\Bigl(\frac{m_e}{2\pi\hbar}\Bigr)^3   \,.
\label{Eq3weibel}
\end{equation}
Here ${\rm Li}_{3/2}$ is a polylogarithm function \cite{Abramowitz72,Lewin81}. Also,
$\beta = 1/[\kappa_{B}(T_{\bot}^2 T_{\parallel})^{1/3}]$, where $T_{\bot}$ and $T_\parallel$ are
related to velocity dispersion in the direction perpendicular and parallel to $z$ axis, respectively.
In the special case when $T_\bot = T_\parallel$, the usual Fermi-Dirac equilibrium is recovered.
The chemical potential is obtained by solving the normalization condition (\ref{Eq3weibel}), yielding,
in particular, $\mu = E_F$ in the limit of zero temperature, where
$E_F = (3\pi^2 n_0)^{2/3}\hbar^2/(2m_e)$ is the Fermi energy.
Also, the Fermi-Dirac distribution $\hat{f}({\bf k})$, where ${\bf k}$ is the appropriated
wave vector in momentum space, is related to the equilibrium Wigner function (\ref{Eq2weibel}) by
$\hat{f}({\bf k}) = (1/2) (2\pi\hbar/m)^3 f_{0}({\bf v})$, with the factor $2$ coming from
spin \cite{Ross60,Arista84}. However, these previous works refer to the cases where there is
no temperature anisotropy. Notice that it has been suggested \cite{Leemans92} that in laser
plasmas the Weibel instability is responsible for further increase of $T_\parallel$ with time.

Inserting (\ref{Eq2weibel}) into (\ref{Eq1weibel}) and integrating over the perpendicular velocity
components, we obtain
\begin{equation}
\omega^2 -  k^2 c^2 - \omega_{pe}^2 \left(1 + \frac{T_\bot}{T_{\parallel}}W_Q\right) = 0 \,,
\label{Eq4weibel}
\end{equation}
where
\begin{equation}
W_Q = \frac{1}{2\sqrt{\pi} H {\rm Li}_{3/2}(-e^{\beta\mu})}\int\frac{d\nu}{\nu-\xi}
\Biggl({\rm Li}_{2}\Bigl\{-\exp\Bigl[-\Bigl(\nu + \frac{H}{2}\Bigl)^2 +
\beta\mu\Bigl]\Bigr\}
- {\rm Li}_{2}\Bigl\{-\exp\Bigl[-\Bigl(\nu - \frac{H}{2}\Bigl)^2 + \beta\mu\Bigl]\Bigr\}\Biggr) \,.
\label{Eq5weibel}
\end{equation}
In (\ref{Eq5weibel}), ${\rm Li}_{2}$ is the dilogarithm function \cite{Abramowitz72,Lewin81},
$H = \hbar k/(m_e v_{\parallel})$ is a characteristic parameter representing the
quantum diffraction effect, $\xi = \omega/(kv_{\parallel})$,
and $\nu = v_{z}/v_{\parallel}$, with $v_\parallel = (2\kappa_{B}T_{\parallel}/m_e)^{1/2}$.
In the simultaneous limit of a small quantum diffraction effect ($H \ll 1$) and
a dilute system ($e^{\beta\mu} \ll 1$), it can be shown that  $W_Q \simeq - 1 - \xi Z(\xi)$,
where $Z$ is the standard plasma dispersion function \cite{Fried61}.
It is important to notice that either (\ref{Eq1weibel}) or (\ref{Eq4weibel}) reproduces the transverse
dielectric function calculated from the random phase approximation for a fully degenerate
quantum plasma \cite{Lindhard54}, in the case of an isotropic system. The simple way to verify
this equivalence is to put $T_\bot = T_{\parallel}$ in (\ref{Eq1weibel}) and then take the
limit of zero temperature, so that $f_0 = 3n_{0}/(4\pi V_{Fe}^3)$ for $|{\bf v}| < V_{Fe}$,
and $f_0 = 0$ otherwise.

\begin{figure}[htb]
\centering
\includegraphics[width=8.5cm]{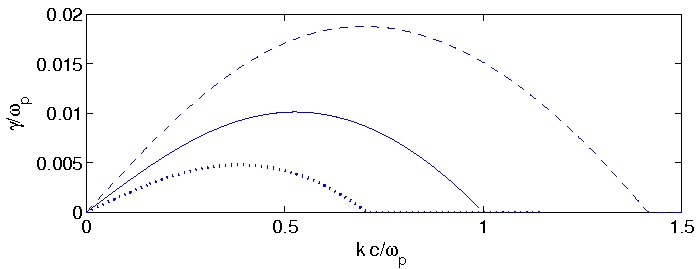}
\caption{The growth rate for the Weibel instability of a dense Fermionic plasma
with $n_0=10^{33}\,\mathrm{m}^{-3}$ ($\omega_{pe}=1.8\times 10^{18}\,\mathrm{s}^{-1}$) and $\beta\mu=5$,
relevant for the next generation inertially compressed material in intense laser-solid density
plasma interaction experiments.  The temperature anisotropies are $T_\perp/T_{||}=3$ (dashed line),
$T_\perp/T_{||}=2$ (solid line) and $T_\perp/T_{||}=1.5$ (dotted line), yielding, respectively,
$T_{||}=3.9\times 10^6\,\mathrm{K}$, $T_{||}=5.2\times 10^6\,\mathrm{K}$
and $T_{||}=6.3\times 10^6\,\mathrm{K}$. After Ref. \cite{Haas09}.}
\label{Fig1weibel}
\end{figure}

\begin{figure}[htb]
\centering
\includegraphics[width=8.5cm]{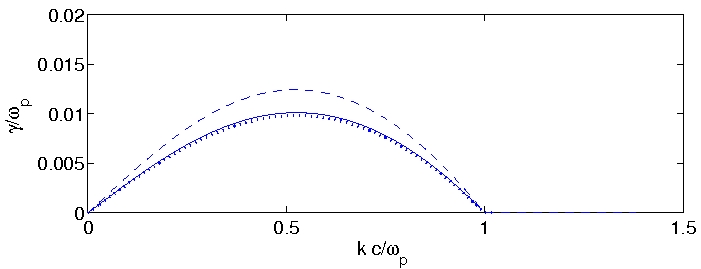}
\caption{The growth rate for the Weibel instability of a dense Fermionic plasma with $n_0=10^{33}\,
\mathrm{m}^{-3}$ ($\omega_{pe}=1.8\times 10^{18}\,\mathrm{s}^{-1}$). Here the temperature anisotropy
is $T_\perp/T_{||}=2$. We used $\beta\mu=1$ (dashed line), $\beta\mu=5$ (solid line) and
$\beta\mu=10$ (dotted line), yielding $T_{||}=1.6\times 10^7\,\mathrm{K}$,
$T_{||}=5.2\times 10^7\,\mathrm{K}$ and $T_{||}=2.6\times 10^6\,\mathrm{K}$, respectively. After Ref. \cite{Haas09}.}
\label{Fig2weibel}
\end{figure}

We next solve our new dispersion relation (\ref{Eq4weibel}) for a set of parameters that are representative
of the next generation laser-solid density plasma interaction experiments.  The normalization
condition (\ref{Eq3weibel}) can also be written as
$-{\rm Li}_{3/2}[-\exp(\beta\mu)]=(4/3\sqrt{\pi})(\beta  E_F)^{3/2}$, which
is formally the same relation holding for isotropic Fermi-Dirac equilibria \cite{Bransden00}.
For a given value on the product $\beta\mu$ and the density, this relation yields the
value $\beta$, from which the temperatures $T_\perp$ and $T_{||}$ can be calculated,
if we know $T_\perp/T_{||}$. Consider only purely growing modes.
From the definition (\ref{Eq5weibel}), one can show that $W_Q\rightarrow -1$ when
$\omega=i\gamma\rightarrow 0$ for a finite wavenumber $k$.
From (\ref{Eq4weibel}) we then obtain the maximum wavenumber for instability as
$k_{\rm max}=(\omega_{pe}/c)\sqrt{T_\perp/T_{||}-1}$. When
$T_\perp/T_{||}\rightarrow 1$, the range of unstable wavenumbers shrinks to
zero. In Figs. \ref{Fig1weibel} and \ref{Fig2weibel}, we have used the
electron number density $n_0=10^{33}\,\mathrm{m}^{-3}$, which can be
obtained in laser-driven compression schemes. The growth rate
for different values on $T_\perp/T_{||}$ is displayed in Fig. \ref{Fig1weibel}. We see
that the maximum unstable wavenumber is $k_{\rm max}=(\omega_{pe}/c)\sqrt{T_\perp/T_{||}-1}$,
as predicted, and that the maximum growth rate occurs at $k\approx k_{\rm max}/2$.
Figure \ref{Fig1weibel} also reveals that the maximum growth rate of the instability is
almost linearly proportional to $T_\perp/T_{||}-1$. In Fig. \ref{Fig2weibel}, we have varied
the product $\beta\mu$, which is a measure of the degeneracy of the quantum plasma.
We see that for $\beta\mu$ larger than $5$, the instability reaches a limiting value,
which is independent of the temperature, while thermal effects start to play an important
role for $\beta\mu$ of the order unity.

\begin{figure}[htb]
\centering
\includegraphics[width=8.5cm]{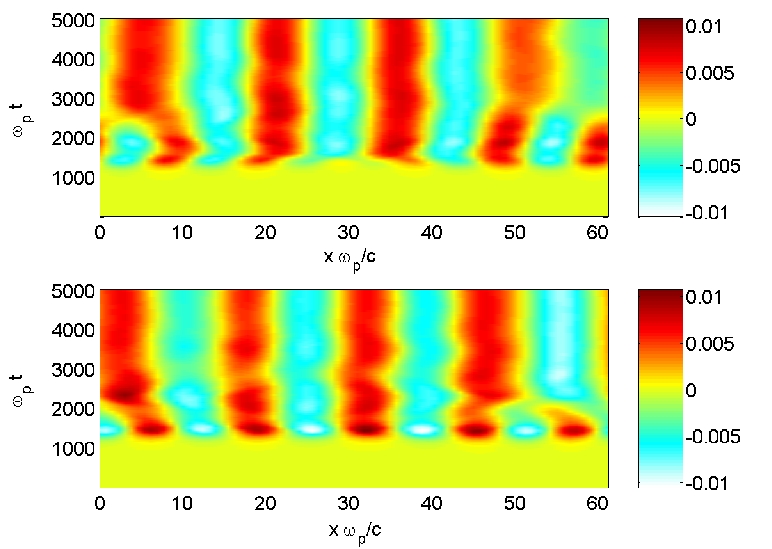}
\caption{The magnetic field components $B_y$ (top panel) and $B_z$
(bottom panel) as a function of space and time, for $\beta\mu=5$ and $T_{\perp}/T_{||}=2$.
The magnetic field has been normalized by $\omega_{pe} m_e/e$.  We see a nonlinear saturation
of the magnetic field components at an amplitude of $\sim 0.01$. After Ref. \cite{Haas09}.}
\label{Fig3weibel}
\end{figure}

\begin{figure}[htb]
\centering
\includegraphics[width=8.5cm]{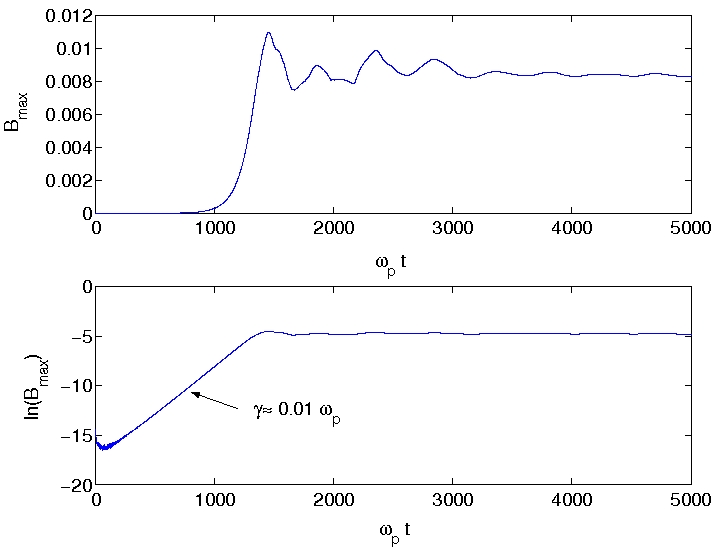}
\caption{The maximum of the magnetic field amplitude, $B=(B_y^2+B_z^2)^{1/2}$, over
the simulation box (top panel), and the logarithm of the magnetic field maximum (bottom panel)
as a function of time, for $T_\perp/T_{||}=2$ and $\beta\mu=5$.  The magnetic field has been
normalized by $\omega_{pe} m_e/e$.  From the logarithmic slope of the magnetic field in the linear regime
one finds $\gamma\approx\Delta {\rm ln}(B_{\rm max})/\Delta t\approx 0.01\,\omega_{p}$. After Ref. \cite{Haas09}.}
\label{Fig4weibel}
\end{figure}

From several numerical solutions of the linear dispersion relation, it was deduced \cite{Haas09} an
approximate scaling law for the instability as
$\gamma_{max}/\omega_{pe}={\rm constant}\times n_0^{1/3}(T_\perp/T_{||}-1)$,
where the constant is approximately $8.5\times 10^{-14}\,\mathrm{s}^{-1}\mathrm{m}$.
Using that $n_0=(2 m_e E_F/\hbar^2)^{3/2}/(3\pi^2) \approx 1.67\times10^{36}(E_F/m_e c^2)^{3/2}$,
we have
\begin{equation}
  \frac{\gamma_{\rm max}}{\omega_{pe}}=0.10\left(\frac{E_F}{m_e c^2}\right)^{1/2}
  \left(\frac{T_\perp}{T_{||}}-1\right),
  \label{Eq7weibel}
\end{equation}
for the maximum growth rate of the Weibel instability in a degenerate Fermi plasma.
This scaling law, where the growth rate depends on the Fermi energy and the temperature anisotropy,
should be compared to that of a classical plasma \cite{Trivelpiece73,Estabrook78}, where the growth rate
depends on the thermal energy and the temperature anisotropy.

For a Maxwellian plasma, it has been found \cite{Davidson72} that the Weibel
instability saturates nonlinearly once the magnetic bounce frequency
$\omega_c=eB/m_e c$ has increased to a value comparable to the linear growth rate.
In order to assess the nonlinear behavior of the Weibel instability for a degenerate plasma, we
have carried out a kinetic simulation of the Wigner-Maxwell system.
We have assumed that the quantum diffraction effect is small, so that the simulation of the
Wigner equation can be approximated by simulations of the Vlasov equation by means of an
electromagnetic Vlasov code \cite{Eliasson07}. As an initial condition for the simulation,
we used the distribution function (\ref{Eq2weibel}). In order to give a seed for any instability,
the plasma density was perturbed with low-frequency fluctuations (random numbers).
The results are displayed in Figs. \ref{Fig3weibel} and \ref{Fig4weibel},
for the parameters $\beta\mu=5$ and $T_\perp/T_{||}=2$,
corresponding to the solid lines in Figs. \ref{Fig1weibel} and \ref{Fig2weibel}.
Figure \ref{Fig3weibel} shows the magnetic field
components as a function of space and time. We see that the magnetic field initially grows,
and saturates to steady state magnetic field fluctuations with an amplitude
of $eB/m_e c \omega_{pe}\approx 0.008$. The maximum amplitude of the magnetic field
over the simulation box as a function of time is shown in Fig. \ref{Fig4weibel}, where we see
that the magnetic field saturates at $eB/m_e \omega_{pe}\approx 0.0082$, while the linear
growth rate of the most unstable mode is $\gamma_{max}/\omega_{pe} \approx 0.009$.
Similar to the classical Maxwellian plasma case \cite{Davidson72},
we can thus estimate the generated magnetic field as
\begin{equation}
B = \frac{m_e c \gamma_{\rm max}}{e},
\end{equation}
for a degenerate Fermi plasma.  For our parameters parameters relevant for intense laser-solid
interaction experiments, we will thus have magnetic fields of the order $10^5\,\mathrm{Tesla}$
(one gigagauss).

\subsection{Dense plasma magnetization by the electromagnetic wave
\label{sec:Magnetization}}

The second example of the dense plasma magnetization can occur in the presence of streaming electrons and
large amplitude electromagnetic waves. Here, the nonstationary ponderomotive force of the
electromagnetic wave would create slowly varying electric fields and currents, which generate d.c. magnetic
fields in dense plasmas.

Let us consider the propagation of the electromagnetic wave with the electric field
${\bf E} ({\bf r}, t) = (1/2) {\bf E}_0 (x, t) \exp(-i \omega t + i k x)$ + c.c., in an unmagnetized
non-relativistic dense plasma with streaming electrons (with the drift velocity $ u \hat {\bf z}$, where $u$ is
the magnitude of the electron drift speed and $\hat {\bf z}$ is the unit vector along the $z$ axis
in a Cartesian coordinate system; typically $u$ is much smaller than the electron
Fermi speed) and immobile ions.  Here, ${\bf E}_0 (x, t)$ is the envelope of the electromagnetic
field at the position ${\bf r}$ and time $t$, and $c. c.$ stands for complex conjugate.
The frequency $\omega$ and the wave vector ${\bf k} =k \hat {\bf x}$, where $\hat {\bf x}$
is the unit vector along the $x$ axis, are related by \cite{Shukla09b}

\begin{equation}
\frac{k^ 2c^ 2}{\omega^ 2} = N = 1- \frac{\omega_{pe}^ 2}{\omega^ 2}-
\frac{k^ 2u^ 2\omega_{pe}^ 2}{\omega^ 2(\omega^ 2-k^ 2 V_{Fe}^ 2- \Omega_q^ 2)},
\label{Eq_M1}
\end{equation}
where $N$ is the index of refraction and $\Omega_q =\hbar k^ 2/2m_e$,

The electromagnetic wave exerts a ponderomotive force ${\bf F}_p
={\bf F}_{ps} + {\bf F}_{pt}$ on the plasma electrons,
where the stationary and non-stationary ponderomotive forces \cite{Karpman76}
are, respectively,

\begin{equation}
{\bf F}_{ps} = \frac{(N-1)}{16 \pi}\nabla |{\bf E}_0|^2,
\label{Eq_M2}
\end{equation}
and

\begin{equation}
{\bf F}_{pt}= \frac{1}{16 \pi}\frac{\bf k}{\omega^ 2}
\frac{\partial[\omega^ 2(N-1)]}{\partial \omega}
\frac{\partial |{\bf E}_0|^ 2}{\partial t}.
\label{Eq_M3}
\end{equation}

The ponderomotive force pushes the electrons locally, and creates the
slowly varying electric field

\begin{equation}
{\bf E}_s =-\nabla \phi_s - \frac{1}{c}\frac{\partial {\bf A}_s}{\partial t}
=\frac{1}{n_0 e}{\bf F}_p,
\label{Eq_M4}
\end{equation}
where the scalar and vector potentials are, respectively,

\begin{equation}
{\bf \phi}_s =-\frac{(N-1)}{16\pi n_0e}|{\bf E}_0|^ 2,
\label{Eq_M5}
\end{equation}
and

\begin{equation}
{\bf A}_s = -\frac{c}{16\pi n_0 e}\frac{{\bf k}} {\omega^ 2}
\frac{\partial[\omega^ 2(N-1)]}{\partial \omega} |{\bf E}_0|^ 2.
\label{Eq_M6}
\end{equation}

The induced slowly varying magnetic field ${\bf B}_s$ is then
${\bf B}_s =\nabla \times {\bf A}_s$.  Noting that

\begin{equation}
\frac{\partial[\omega^ 2(N-1)]}{\partial \omega}
= \frac{2 \omega k^ 2 u^ 2 \omega_{pe}^ 2}
{(\omega^ 2- k^ 2 V_{Fe}^ 2 -\Omega_q^ 2)^ 2},
\label{Eq_M7}
\end{equation}
we can express the magnitude of the magnetic field as

\begin{equation}
  |{\bf B}_s| =  \frac{ e c  k^ 3 u^ 2|{\bf E}_0|^2}
{2 m_e L \omega (\omega^ 2- \Omega^ 2)^2},
  \label{Eq_M8}
\end{equation}
where $L$ is scale length of the envelope $|{\bf E}_0|^2$ and $\Omega =
(k^ 2V_{Fe}^ 2+ \Omega_q^ 2)^ {1/2} \equiv (\hbar k^2/2m_e)[1+ 4(3\pi^ 2 n_0)^ {2/3}/k^ 2]^{1/2} $.
We note from (\ref{Eq_M8}) that the magnetic field strength is proportional to $u^ 2$, and
attains a large value when $\omega \sim \Omega$. The electron gyrofrequency $\Omega_c$ is

\begin{equation}
  \Omega_c =  \frac {e|{\bf B}_s|}{m_e c}
  = \frac{k^ 3 V_0^2 u^ 2 \omega}{2 L(\omega^ 2-\Omega^ 2)^ 2},
  \label{Eq_M9}
\end{equation}
where $V_0 =e|{\bf E}_0|/m_e \omega$ is the electron quiver velocity in the electromagnetic field.

\section{Dynamics of electromagnetic waves in dense plasmas\label{emwaveequation}}

We here consider various electromagnetic (EM) wave modes and their nonlinear interaction
in dense plasmas. As examples, we will discuss spin waves, which occur in a
magnetized plasma due to the spin-1/2 effect of electrons and positrons.
We also consider nonlinear interactions between finite amplitude electromagnetic
and electrostatic waves in dense plasmas. We focus on the underlying physics of
stimulated scattering instabilities and the wave localization due to the parametric
interactions involving the radiation pressure \cite{Shukla86}.

\subsection{Electromagnetic spin waves in magnetized plasmas}

In the presence of an external magnetic field, the quantum description of the linear kinetics of
a dense collisionless plasma is complicated due to quantization of the gyromagnetic motion and the
inclusion of the electron and positron spin-$1/2$. In the past, many authors
\cite{Oberman63,Kelley64,Benford69,Kuzelev99,Melrose02} investigated the high-frequency conductivity, longitudinal
and transverse dielectric responses in dense magnetized plasmas. The propagation characteristics
of high-frequency electromagnetic waves in quantum magnetoplasmas is different from that in
a classical magnetoactive plasma \cite{Ginzburg60}.

Recently, Oraevsky et al.  \cite{Semikoz03} found a new electromagnetic spin wave whose electric
field is parallel to $\hat {\bf z} B_0$, and which propagates across $\hat {\bf z}$, where
$\hat {\bf z}$ is the unit vector along the $z$ axis in a Cartesian coordinate system and $B_0$ is
the strength of the ambient magnetic field. The spin wave accompany the magnetization current
due to spinning electron motion and the wave frequency is obtained from \cite{Semikoz03}

\begin{equation}
\frac{k_\perp^2c^2}{\omega^2} = 1- \frac{\omega_{pe}^2}{\omega(\omega + i \nu_e)}
+ 2\pi \mu_B^2 \frac{k_\perp^2 c^2n_0}{\omega^2 \epsilon_F}\frac{\omega_{ce}}{\omega -\omega_{ce}},
\label{spin1}
\end{equation}
where $k_\perp$ is the perpendicular component of the wave vector ${\bf k} = \hat {\bf x} k_\perp$,
$\hat {\bf x}$ is the unit vector transverse to $\hat {\bf z}$, $\nu_e$ is the electron collision
frequency, nd $\epsilon_F \sim m_e c^2$. The third term in the right-hand side of (102) represents
the electron spin-magnetic resonance at the electron gyrofrequency $\omega_{ce} =eB_0/m_e c$.

Assuming that $k_\perp^2 c^2 + \omega_{pe}^2 \neq \omega_{ce}^2$, we obtain from (102)

\begin{equation}
\omega \simeq \omega_{ce}\left[ 1+ \frac{2\pi \mu_B^2n_0}{\epsilon_F}
\frac{k_\perp^2 c^2}{k_\perp^2c^2 + \omega_{pe}^2 \left(1-i\nu_e/\omega\right)
-\omega_{ce}^2} \right].
\end{equation}
For $\nu_e \ll \omega \sim \omega_{ce}$, the damping rate of the spin mode is

\begin{equation}
{\rm Im} \omega \simeq \frac{\nu_e \omega_{pe}^2k_\perp^2 c^2\mu_B^2B_0^2}
{ \epsilon_F m_e c^2\omega_{ce}^2(k_\perp^2 c^2 +\omega_{pe}^2-\omega_{ce}^2)^2}.
\end{equation}

The ponderomotive force of the spin wave can create compressional magnetic field
perturbation due to an inverse Cotton-Mouton/Faraday effect \cite{Abdul88}.

\subsection{Nonlinearly coupled EM waves}

Finite amplitude electromagnetic waves in quantum magnetoplasmas interact nonlinearly among themselves.
In this subsection, we shall use the generalized Q-MHD equations to obtain compact nonlinear equations
for the electron magnetohydrodynamic (EMHD) and the Hall-MHD plasmas, and show how does the density,
the fluid velocity, and magnetic field perturbations are coupled in a non-trivial manner.

The governing nonlinear equations for the electromagnetic waves in dense magnetoplasmas
are the quantum magnetohydrodynamic equations composed of the continuity equation

\begin{equation}
\frac{\partial n_{e,i}}{\partial t} + \nabla \cdot (n_{e,i} {\bf u}_{e,i}) = 0,
\label{Eq_EM1}
\end{equation}
the electron and ion momentum equations, respectively,


\begin{equation}
n_e m_e \left(\frac{\partial}{\partial t}
+ {\bf u}_e \cdot \nabla\right) {\bf u}_e
= - n_e e  \left({\bf E} + \frac{1}{c}
{\bf u}_e \times {\bf B} \right) - \nabla P_e + {\bf F}_{Qe},
\label{Eq_EM2}
\end{equation}
\begin{equation}
n_i m_i \left(\frac{\partial}{\partial t}
+ {\bf u}_i \cdot \nabla\right) {\bf u}_i
= Z_i e n_i   \left({\bf E} + \frac{1}{c}
{\bf u}_i \times {\bf B} \right),
\label{Eq_EM5}
\end{equation}
the Faraday law

\begin{equation}
c \nabla \times {\bf E} =- \frac{\partial {\bf B}}{\partial t},
\label{Eq_EM3}
\end{equation}
the Maxwell equation including the magnetization spin current

\begin{equation}
\nabla \times {\bf B} = \frac{4\pi }{c}\left({\bf J}_p + {\bf J}_m\right)
+ \frac{1}{c}\frac{\partial {\bf E}}{\partial t},
\label{Eq_EM4}
\end{equation}
where the pressure for non-relativistic degenerate electrons reads \cite{Eliezer05}

\begin{equation}
P_e = \frac{4 e B (2 m_e)^{1/2} E_F^{3/2}} {3 (2\pi)^2\hbar^2 c}
\left[1+2\sum_{n_L=1}^{n_{max}}\left(1-\frac{n_L\hbar \omega_{ce}} {E_{F}}\right)^{3/2}\right],
\label{Eq_EM6}
\end{equation}
where $n_L =0,\,1,\,2,\,\ldots,\,n_{max}$, and the value of $n_{max}$ is fixed by the largest integer $n_L$ that
satisfies $E_F-n_L\hbar\omega_{ce}\geq 0$.


The sum of the quantum Bohm and intrinsic angular momentum spin forces is

\begin{equation}
{\bf F}_{Qe} =  \nabla \left(\frac{\nabla^ 2\sqrt{n_e}}{\sqrt{n_e}}\right)
-\frac{n_e \mu_B^2}{k_B T_{Fe}} \nabla B.
\label{Eq_EM7}
\end{equation}

In Eqs. (\ref{Eq_EM1})--(\ref{Eq_EM7}), $n_j$ is the number density of the particle species $j$ ($j$
equals $e$ for the electrons, and $i$ for the ions), ${\bf u}_j$ is the particle fluid velocity,
and $B=|{\bf B}|$. We have denoted the plasma current density ${\bf J}_p = - n_e e {\bf u}_e + Z_i
n_i e {\bf u}_i $ and the electron magnetization spin current density ${\bf J}_m = \nabla \times {\bf M}$,
where the magnetization for dynamics on a time scale much slower than the spin precession frequency
for an electron Fermi gas reads \cite{Kittel} ${\bf M} = (n_e \mu_B^2/k_B T_{Fe}) \hat {\bf B}$.

\subsubsection{Nonlinear EMHD}

First, we present the generalized nonlinear EMHD equations for a dense magnetoplasma.
Here the ions form the neutralizing background. The wave phenomena in the EMHD plasma
will occur on a time scale much shorter than the ion plasma and ion gyroperiods.
In equilibrium,  we have

\begin{equation}
n_{e0} = Z_i n_{i0} \equiv n_0.
\label{Eq_EM9}
\end{equation}

The relevant nonlinear EMHD equations are
\begin{equation}
\frac{\partial n_e}{\partial t} + \nabla \cdot (n_e {\bf u}_e) =0,
\label{Eq_EM10}
\end{equation}
the electron momentum equation (\ref{Eq_EM2}), Faraday's law (\ref{Eq_EM3}), and the
electron fluid velocity given by

\begin{equation}
{\bf u}_e = \frac{{\bf J}_m}{e n_e}-\frac{c (\nabla \times {\bf B})}{4 \pi e n_e}
+ \frac{1}{e n_e}\frac{\partial {\bf E}}{\partial t}.
\label{Eq_EM11}
\end{equation}
We observe that the quantum tunneling and spin forces play an important role if there are
slight electron density and magnetic field inhomogeneities in dense plasmas. The nonlinear
EMHD equations are useful for studying collective electron dynamics in metallic and
semiconductor nanostructures \cite{r1a}.

\subsubsection{Nonlinear Hall-MHD}

Second, we derive the  modified nonlinear Hall-MHD equations in a dense electron-ion
plasma. The Hall-MHD equations shall deal with the wave phenomena on a time scale larger
than the electron gyroperiod. The relevant nonlinear Hall-MHD equations are the
electron and ion continuity equations, the inertialess electron momentum equation


\begin{equation}
{\bf E} + \frac{1}{c} {\bf u}_e \times {\bf B} - \frac{\nabla P_e}{n_e e}
- \frac{{\bf F}_{Qe}}{n_e e} =0,
\label{Eq_EM12}
\end{equation}
Faraday's law (\ref{Eq_EM3}), the ion momentum equation

\begin{equation}
n_i  m_i \left(\frac{\partial}{\partial t}
+ {\bf u}_i \cdot \nabla \right) {\bf u}_i
= Z_i e n_i   \left({\bf E} + \frac{1}{c}
{\bf u}_i \times {\bf B} \right),
\label{Eq_EM13}
\end{equation}
and the electron fluid velocity given by

\begin{equation}
{\bf u}_e = {\bf u}_i + \frac{{\bf J}_m}{e n_e}
-\frac{c (\nabla \times {\bf B})}{4 \pi e n_e},
\label{Eq_EM15}
\end{equation}
where we have neglected the displacement current, since the Hall-MHD plasma deals with
electromagnetic waves whose phase velocity is much smaller than the speed of light in
vacuum.

We now eliminate the electric field from (\ref{Eq_EM13}) by using (\ref{Eq_EM12}),
obtaining

\begin{equation}
n_i m_i \left(\frac{\partial}{\partial t}
+ {\bf u}_i \cdot \nabla \right) {\bf u}_i
= Z_i e n_i   \left[\frac{1}{c}
({\bf u}_i-{\bf u}_e) \times {\bf B} - \frac{\nabla P_e}{n_e e}
+ \frac{{\bf F}_{Qe}}{n_e e} \right] .
\label{Eq_EM16}
\end{equation}

Furthermore, eliminating ${\bf u}_e$ from (\ref{Eq_EM16}) by using (\ref{Eq_EM15}),
we have

\begin{equation}
n_i m_i \left(\frac{\partial}{\partial t}
+ {\bf u}_i \cdot \nabla\right) {\bf u}_i
= Z_i e n_i   \left\{\frac{1}{c}
\left[- \frac{{\bf J}_m}{e n_e} +
\frac{c (\nabla \times {\bf B})}{4\pi} \right] \times {\bf B}
- \frac{\nabla P_e}{n_e e}
+ \frac{{\bf F}_{Qe}}{n_e e} \right\},
\label{Eq_EM17}
\end{equation}
with the quasi-neutrality condition $n_e = Z_i n_i$.

Finally, by using (\ref{Eq_EM12}) we can eliminate ${\bf E}$ from (\ref{Eq_EM2}),
obtaining

\begin{equation}
\frac{\partial {\bf B}}{\partial t} = \nabla \times \left[ ({\bf u}_i \times {\bf B})
+ \frac{{\bf J}_m \times {\bf B}}{e Z_i n_i}
-\frac{c}{4\pi} \left(\nabla \times {\bf B}\right)\times {\bf B} \right].
\label{Eq_EM18}
\end{equation}

The ion continuity equation, Eqs. (\ref{Eq_EM17}) and (\ref{Eq_EM18}), together with (\ref{Eq_EM13})
and $Z_i n_{i1} = n_{e1}$, where $n_{e1,i1} \ll n_{e0,i0}$ are the desired generalized nonlinear
equations for the low-frequency (in comparison with the electron gyrofrequency), low-phase velocity
(in comparison with the speed of light in vacuum) in a Hall-MHD dense plasma.  They can be used
to investigate the multi-dimensional linear and nonlinear waves (e.g. magnetosonic solitons \cite{MES07}),
as well as nanostructures and turbulences \cite{Shaikh09} in dense quantum magnetoplasmas.

\subsection{Stimulated scattering instabilities}

Nonlinear interactions between the high-frequency EM waves and low-frequency electrostatic
waves give rise to stimulated scattering instabilities in classical
plasmas \cite{Drake74,Yu74,Shukla78,Stenflo90}.
There also exists possibility of exciting plasma waves by the high-frequency EM waves
in quantum plasmas due to the parametric instabilities \cite{Sagdeev69}. The governing equations
for the high-frequency electromagnetic waves \cite{ShuklaStenflo06} and the radiation pressure
driven modified Langmuir and ion-acoustic oscillations in an unmagnetized quantum plasma are,
respectively,

\begin{equation}
\left(\frac{\partial^2}{\partial t^2} -c^2 \nabla^2 + \omega_{pe}^2\right) {\bf A} + \omega_{pe}^2
\frac{n_1}{n_0} {\bf A} \approx 0,
\label{QEM9}
\end{equation}

\begin{equation}
\left(\frac{\partial^2}{\partial t^2} + \omega_{pe}^2 - \frac{3}{5}V_{Fe}^2 \nabla^2
+ \frac{\hbar^2}{4m_e^2}
\nabla^4\right) \frac{n_1}{n_0} = \frac{q_e^2}{2m_e^2c^2}\nabla^2|{\bf A}|^2,
\label{QEM10}
\end{equation}
and

\begin{equation}
\left(\frac{\partial^2}{\partial t^2} -\frac {m_e}{m_i}V_{Fe}^2 \nabla^2 + \frac{\hbar^2}{4 m_e m_i}
\nabla^4\right) \frac{n_1}{n_0} = \frac{q_e^2}{2m_e m_i c^2}\nabla^2 |{\bf A}|^2,
\label{QEM11}
\end{equation}
where ${\bf A}$ is the vector potential of the high-frequency electromagnetic wave,
and $n_1$ is the electron density perturbation of the low-frequency oscillations
(viz. the modified EPOs and ion waves).

Combining Eqs. (\ref{QEM9})-(\ref{QEM11}) we thus simply obtain the nonlinear dispersion relations

\begin{equation}
\omega^2 - \Omega_R^2
=- \frac{q_e^2 \omega_{pe}^2 k^2|{\bf A}_0|^2}{2m_e^2c^2}\sum_{\pm}\frac{1}{D_\pm},
\label{QEM12}
\end{equation}
and

\begin{equation}
\omega^2 - \Omega_B^2
= - \frac{q_e^2\omega_{pe}^2k^2 |{\bf A}_0|^2}{2m_em_ic^2}\sum_{\pm}\frac{1}{D_\pm},
\label{QEM13}
\end{equation}
where $\Omega_R =(\omega_{pe}^2+3k^2V_{Fe}^2/5+ \hbar^2k^4/4m_e^2)^{1/2}$ and
$\Omega_B =(k^2C_{Fs}^2+ \hbar^2 k^4/m_e m_i)^{1/2}$.

Equations (126) and (127) admit stimulated Raman, stimulated Brillouin and modulational instabilities
of the electromagnetic pump (with the amplitude ${\bf A}_0$ and the frequency $\omega_0
=(k_0^2c^2+\omega_{pe}^2)^{1/2}$) in dense quantum plasmas.  We have denoted $D_\pm
= \pm 2 \omega_0(\omega - {\bf k}\cdot {\bf V}_g \mp \delta)$,
where ${\bf V}_g = c^2 {\bf k}_0/\omega_0$ is the group velocity of the
EM pump wave frequency, and $\delta= k^2c^2/2\omega_0$ is the small frequency
shift arising from the nonlinear interaction of the pump with the electrostatic
perturbations $(\omega, {\bf k})$ in our quantum plasma. For stimulated Raman and Brillouin
scatterings, the EM pump wave is scattered off the resonant electron plasma wave and resonant
ion wave, respectively, while for the modulational instability the electron and ion plasma
oscillations are off resonant, so are the EM sidebands.

For three-wave decay interactions, we suppose that $D_- =0$ and $D_+ \neq 0$. Thus, we ignore
$D_+$ from (126) and (127). Letting $\omega =\Omega_R + i \gamma_R$ and $\omega = \Omega_B + i \gamma_B$
in (126) and (127), respectively, and assuming $\omega-{\bf k}\cdot {\bf V}_g + \delta \equiv i \gamma_{R,B}$,
we obtain the the growth rates for stimulated Raman and Brillouin scattering
instabilities (denoted by the subscripts $R$ and $B$, respectively,

\begin{equation}
\gamma_R =\frac{\omega_{pe}ek |{\bf A}_0|^2}{2\sqrt{2 \omega_0 \Omega_R} m_e c},
\label{growthR}
\end{equation}
and
\begin{equation}
\gamma_B =\frac{\omega_{pe}ek |{\bf A}_0|^2}{2\sqrt{2 \omega_0 \Omega_B  m_e m_i} c},
\label{growthB}
\end{equation}
where $|{\bf k}\cdot {\bf V}_g -\delta| \sim \Omega_R, \Omega_B$. We note that the
growth rates for stimulated Raman and Brillouin scattering instabilities are
inversely proportional to the square roots of $\Omega_R$ and $\Omega_B$, which
depend on the quantum parameters.  Stimulated Raman and Brillouin scattering instabilities
of the EM wave off the ES waves should provide invaluable information regarding
the density fluctuations and equation of states that might exist in dense quantum plasmas.

The quantum corrected 3D Zakharov equations \cite{Zakharov72} have been derived by Haas \cite{Haas07}
and Haas and Shukla \cite{HaasShukla09},  who demonstrated that the dispersive effects associated with
quantum corrections can prevent the collapse of localized Langmuir envelope electric fields in both
two and three spatial dimensions.

\subsection{Self-trapped EM waves in a quantum hole}

Nonlinear interactions between large amplitude electromagnetic waves and electrostatic plasma
waves can produce nonlinear nanostructures composed of a density cavity that traps the
electromagnetic wave envelope. Here we demonstrate trapping of intense electromagnetic waves into a
finite amplitude density hole \cite{Shukla07} arising at the scale-size of the order of the
electron skin depth $c/\omega_{pe}$).

A powerful circularly polarized electromagnetic (CPEM) plane wave interacting nonlinearly with
the EPOs would generate an envelope of the CPEM vector potential
${\bf A}_\perp=A_\perp(\hat{\bf x}+i\hat{\bf y}) \exp(-i\omega_0t+ik_0 z)$, which shall obey
the nonlinear Schr\"odinger equation \cite{Shukla86}
\begin{equation}
2i\Omega_0\left(\frac{\partial}{\partial t} +V_g\frac{\partial}{\partial z}\right) A_\perp
+\frac{\partial^2A_\perp}{\partial z^2}
-\left(\frac{|\psi|^2}{\gamma}-1\right)A_\perp=0,
\label{Eq9}
\end{equation}
where the electron wave function $\psi$ and the scalar potential are governed by, respectively,
\begin{equation}
 i H_e \frac{\partial \psi}{\partial t}
 +\frac{H_e^2}{2}\frac{\partial^2\psi}{\partial z^2}+(\phi-\gamma+1)\psi=0,
 \label{Eq10}
\end{equation}
and
\begin{equation}
  \frac{\partial^2\phi}{\partial z^2}=|\psi|^2-1,
  \label{Eq11}
\end{equation}
where the electron number density is defined as is given by the $|\psi|^2$ term.
Here $\Omega_0$ represents the CPEM wave frequency, $V_g$ is the $x$ component of the
group velocity of the CPEM wave, $H_e$ is a quantum coupling parameter, and
$\gamma=(1+|{A}_\perp|^2)^{1/2}$ is the relativistic gamma factor due to the electron
quiver velocity in the CPEM wave fields. The details of normalization of variables is
given in Ref. \cite{Shukla07}. The nonlinear coupling between intense CPEM waves and
EPOs comes about due to the nonlinear current density, which is represented by
the term $|\psi|^2 {A}_\perp/\gamma$ in Eq. (\ref{Eq9}). In Eq. (\ref{Eq10}), $1-\gamma$
is the relativistic ponderomotive potential \cite{Shukla86}, which arises due to
the cross-coupling between the CPEM wave-induced electron quiver velocity and the CPEM wave
magnetic field.

The effect of quantum dispersion on localized electromagnetic pulses can be studied by considering
a steady state structure moving with a constant speed $V_g$. Inserting the ansatz
$A_\perp= W(\xi)\exp(-i\Omega t)$, $\psi= P(\xi)\exp(ikx-i\omega t)$ and $\phi=\phi(\xi)$ into
Eqs.~(\ref{Eq9})--(\ref{Eq11}), where $\xi=z-V_g t$, $k=V_g/H_e$ and
$\omega=V_g^2/2H_e$, and where $W(\xi)$ and $P(\xi)$ are real, one
obtains from (\ref{Eq9})-(\ref{Eq11}) the coupled system of equations
\begin{equation}
  \frac{\partial^2 W}{\partial \xi^2}+\left(\lambda-\frac{P^2}{\gamma}+1\right)W=0,
  \label{Eq13}
\end{equation}
\begin{equation}
  \frac{H_e^2}{2}\frac{\partial^2 P}{\partial \xi^2}+(\phi-\gamma+1)P=0,
  \label{Eq14}
\end{equation}
where $\gamma=(1+W^2)^{1/2}$, and
\begin{equation}
  \frac{\partial^2 \phi}{\partial \xi^2}=P^2-1,
  \label{Eq15}
\end{equation}
with the boundary conditions $W=\Phi=0$ and $P^2=1$ at
$|\xi|=\infty$. In Eq. (\ref{Eq13}), $\lambda=2\Omega_0\Omega$
represents a nonlinear frequency shift of the CPEM wave. In the limit
$H_e\rightarrow 0$, one has from (\ref{Eq14}) $\phi= \gamma -1$, where
$P \neq 0$, and one recovers the classical (non-quantum) case of the
relativistic solitary waves in a cold plasma \cite{Marburger75}.
The system of equations (\ref{Eq13})--(\ref{Eq15}) admits a
Hamiltonian
\begin{eqnarray}
Q_H=\frac{1}{2}\left(\frac{\partial W}{\partial\xi}\right)^2+\frac{H_e^2}{2}
\left(\frac{\partial P}{\partial\xi}\right)^2-\frac{1}{2}\left(\frac{\partial\phi}{\partial\xi}\right)^2
+\frac{1}{2}(\lambda+1)W^2+P^2-\gamma P^2+\phi P^2-\phi=0,
\label{Eq16}
\end{eqnarray}
where the boundary conditions $\partial/\partial\xi=0$, $W=\phi=0$ and $|P|=1$
at $|\xi|=\infty$ have been used.

\begin{figure}[htb]
\centering
\includegraphics[width=8cm]{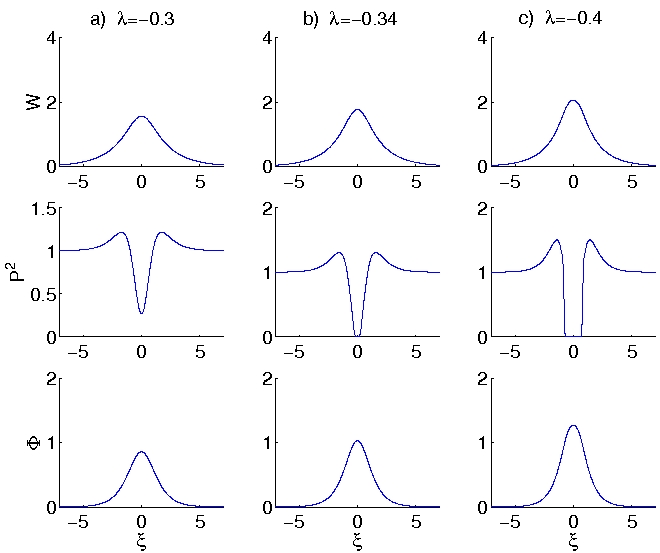}
\caption{The profiles of the CPEM vector potential $W$ (top row), the
electron number density $P^2$ (middle row) and the scalar potential $\Phi$
(bottom row) for $\lambda=-0.3$ (left column), $\lambda=-0.34$
(middle column) and $\lambda=-0.4$ (right column), with $H_e=0.002$.
After Ref.~\cite{Shukla07}.}
\label{Fig6}
\end{figure}

\begin{figure}[htb]
\centering
\includegraphics[width=8cm]{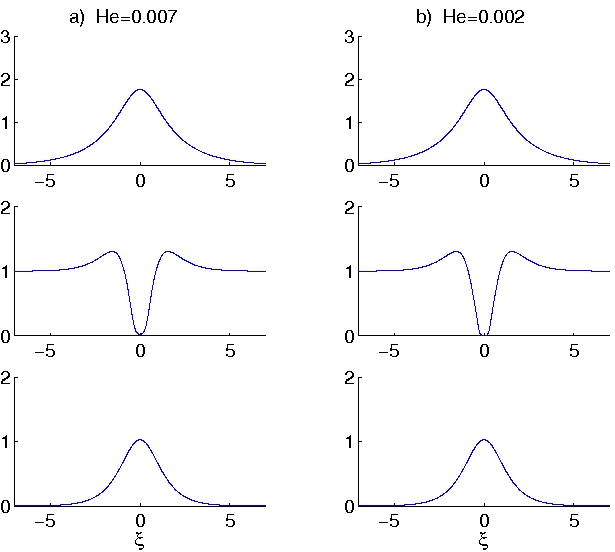}
\caption{The profiles of the CPEM vector potential $W$ (top row), the
electron number density $P^2$ (middle row) and the scalar potential $\Phi$
(bottom row) for $H_e=0.007$ (left column) and $H_e=0.002$ (right column), with
$\lambda=-0.34$. After Ref.~\cite{Shukla07}.}
\label{Fig7}
\end{figure}

\begin{figure}[htb]
\centering
\includegraphics[width=8cm]{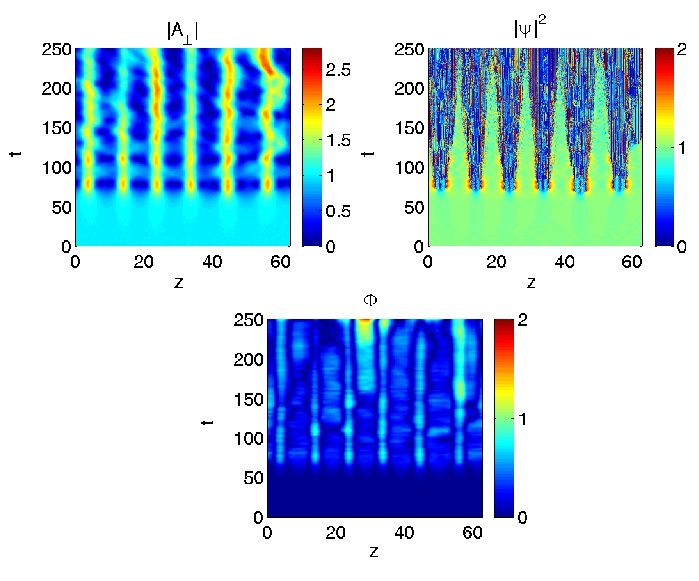}
\caption{The dynamics of the CPEM vector potential ${A}_\perp$ and the electron number
density $|\psi|^2$ (upper panels) and of the electrostatic potential $\Phi$ (lower panel)
for $H_e=0.002$. After Ref.~\cite{Shukla07}.}
\label{Fig8}
\end{figure}

\begin{figure}[htb]
\centering
\includegraphics[width=8cm]{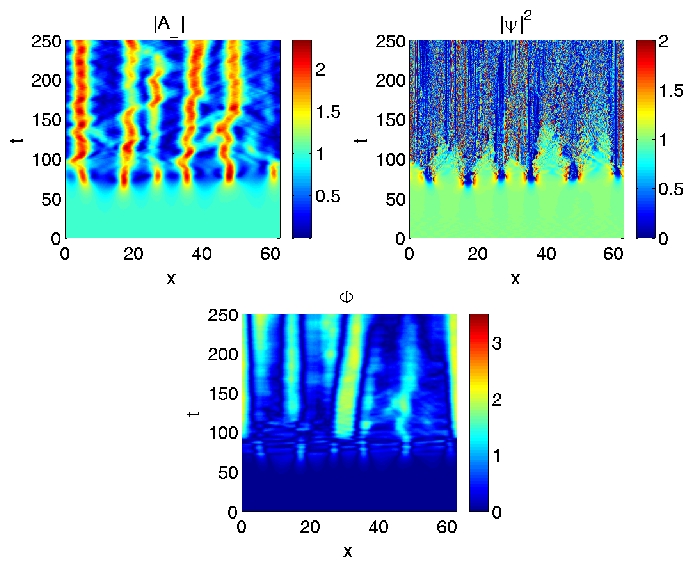}
\caption{
The dynamics of the CPEM vector potential ${A}_\perp$ and the electron number density $|\psi|^2$
(upper panels) and the electrostatic potential $\phi$ (lower panel) for $H_e=0.007$.
After Ref.~\cite{Shukla07}.
}
\label{Fig9}
\end{figure}

Numerical solutions of the quasi-stationary system (\ref{Eq13})--(\ref{Eq15})
are presented  Figs. \ref{Fig6} and \ref{Fig7}, while time-dependent
solutions of Eqs.~(\ref{Eq9})--(\ref{Eq11}) are shown in Figs. \ref{Fig8} and \ref{Fig9}.
Here parameters were used that are representative of the
next generation laser-based plasma compression (LBPC) schemes
\cite{Malkin07,Azechi91,Azechi06}. The formula \cite{Shukla86} $eA_\perp/mc^2 =6 \times
10^{-10} \lambda_s \sqrt{I}$ will determine the normalized vector
potential, provided that the CPEM wavelength $\lambda_s$ (in
microns) and intensity $I$ (in W/cm$^2$) are known. It is expected
that in LBPC schemes, the electron number density $n_0$ may reach
$10^{27}$ cm$^{-3}$ and beyond, and the peak values of
$eA_\perp/mc^2$ may be in the range 1-2 (e.g. for focused EM pulses
with  $\lambda_s \sim 0.15$ nm and $I \sim 5 \times 10^{27}$
W/cm$^{2}$). For $\omega_{pe} =1.76 \times 10^{18}$
s$^{-1}$, one has $\hbar \omega_{pe} =1.76 \times 10 ^{-9}$ erg and
$H_e =0.002$, since $mc^2 =8.1 \times 10^{-7}$ erg. The electron
skin depth $\lambda_e \sim 1.7$ {\AA}. On the other hand,
a higher value of $H_e=0.007$ is achieved for $\omega_{pe} =5.64 \times 10^{18}$
s$^{-1}$. Thus, our numerical solutions below, based on these two values of $H_e$,
have focused on  scenarios that are relevant for the next generation intense
laser-solid density plasma interaction experiments \cite{Malkin07}.

Figures \ref{Fig6} and \ref{Fig7} show numerical solutions of
Eqs. (\ref{Eq13})--(\ref{Eq15}) for several values of $H_e$.
The nonlinear boundary value problem was solved with the
boundary conditions $W=\phi=0$ and $P=1$ at the boundaries at
$\xi=\pm 10$. One can see that the solitary envelope pulse
is composed of a single maximum of the localized vector potential
$W$ and a local depletion of the electron density $P^2$, and a
localized positive potential $\phi$ at the center of the solitary
pulse. The latter has a continuous spectrum in $\lambda$, where
larger values of negative $\lambda$ are associated with larger
amplitude solitary EM pulses. At the center of the solitary EM
pulse, the electron density is partially depleted, as in panels a)
of Fig. \ref{Fig6}, and for larger amplitudes of the EM waves one has a
stronger depletion of the electron density, as shown in panels b)
and c) of Fig. \ref{Fig6}. For cases where the electron density goes to
almost zero in the classical case \cite{Marburger75}, one important
quantum effect is that the electrons can tunnel into the depleted
region. This is seen in Fig. \ref{Fig7}, where the electron density remains
nonzero for the larger value of $H_e$ in panels a), while the density
shrinks to zero for the smaller value of $H_e$ in panel b).

Figures \ref{Fig8} and \ref{Fig9} show numerical simulation results of
Eqs.~(\ref{Eq9})--(\ref{Eq11}), in order to investigate the quantum diffraction effects on the
dynamics of localized CPEM wavepackets. Here the long-wavelength limit
$\omega_0\approx 1$ and $V_g\approx 0$ was considered. In the initial conditions,
an EM pump with a constant amplitude $A_\perp=A_0=1$ and a
uniform plasma density $\psi=1$ was used, together with a small amplitude noise (random
numbers) of order $10^{-2}$ added to ${A}_\perp$ to give a seeding any instability.
The numerical results are displayed in Figs. \ref{Fig8}
and \ref{Fig9} for $H_e=0.002$ and $H_e=0.007$, respectively.
In both cases, one can see an initial linear growth phase and a wave collapse at
$t\approx 70$, in which almost all the CPEM wave energy is contracted into a few well
separated localized CPEM wave pipes. These are characterized by a large
bell-shaped amplitude of the CPEM wave, an almost complete depletion of
the electron number density at the center of the CPEM wavepacket, and
a large-amplitude positive electrostatic potential.
Comparing Fig. \ref{Fig8} with Fig. \ref{Fig9}, one can
see that there is a more complex dynamics in the interaction between
the CPEM wavepackets for the larger $H_e=0.007$, shown in Fig. \ref{Fig9}, in comparison
with $H_e=0.002$, shown in Fig. \ref{Fig8}, where the wavepackets
are almost stationary when they are fully developed.

\section{Summary and perspectives\label{sec:conclusions}}

In this paper, we have presented our up-to-date theoretical knowledge of nonlinear physics
of non-relativistic quantum plasmas. We started with nonlinear quantum models that describe
the physics of localized excitations in different areas of physics. We then moved
to discussing the well known electron and ion plasma wave spectra in dense quantum plasmas,
and presented nonlinear models for treating nonlinear interactions among finite amplitude
plasma waves at nanoscales. As examples, we demonstrated the existence of localized nonlinear
EPOs and localized ion waves in dense quantum plasmas. For electrostatic EPOs, the electron dynamics
is governed by a pair composed of nonlinear Schr\"odinger and Poisson (NLSP) equations.
We stress that the nonlinear Schr\"odinger equation governing the spatio-temporal
evolution of the electron wave function in the presence of self-consistent electrostatic
potential in dense plasmas has been obtained from the quantum electron momentum equation by
introducing the eikonal representation as a mathematical tool for understanding the
complex electron plasma wave interactions at nanoscales. Such a description also appears
in the context of the nonlinear electron  dynamics in thin metal films \cite{r1a}.
Our NLSP equations admit a set of conserved quantities, e.g. the total number of electrons,
the electron momentum, the electron angular momentum, and the electron energy. We have
found that the NLSP equations admit quasi-stationary, localized structures in the form of
one-dimensional quantized dark solitons and two-dimensional quantized vortices.
These nanostructures are associated with a local depletion of the electron density
associated with positive electrostatic potential, and are parameterised by the quantum coupling
parameter only. In the two-dimensional geometry, there exist a class of vortices of different
excited states (charge states) associated with a complete depletion of the electron density and an
associated positive potential. Numerical simulations of the time-dependent NLSP equations
demonstrate the stability of stable dark solitons in one space dimension with an amplitude
consistent with the one found from the time-independent solutions. In two-space-dimensions,
the dark solitons of the first excited state were found to be stable and the preferred
nonlinear state was in the form of  vortex pairs of different polarities. One-dimensional
dark solitons and singly charge two-dimensional vortices are thus long-lived nonlinear
structures in quantum plasmas.

Similar to the Bernstein-Green-Kruskal modes \cite{EliassonShukla06}, we accounted for the
trapping of electrons in the electrostatic wave potential and studied numerically the deformation
of the equilibrium Fermi-Dirac distribution function and the subsequent emergence of
localized phase space kinetic structures. Such an investigation was carried out by
using the time-dependent Wigner and Poisson equations. Furthermore, we presented
two mechanisms for the generation of magnetic fields in quantum plasmas. They are
associated with the quantum Weibel instability in the presence of equilibrium anisotropic
Fermi-Dirac electron distribution function, as well as with the the non-stationary ponderomotive
force of a large amplitude electromagnetic waves in quantum plasmas with streaming electrons.
Spontaneously generated magnetic fields can affect the linear and nonlinear propagation of
both the electrostatic and electromagnetic waves in quantum magnetoplasmas \cite{ShuklaAli06}.
The quantum corrections produce dispersion at short scales for the electrostatic upper-hybrid,
lower-hybrid, and ion-cyclotron waves, while the quantum Bohm force and the electron$-1/2$ spin
effects introduce new features to the elliptically polarized extraordinary electromagnetic mode
\cite{Shukla07a}. Furthermore, the electron-$1|2$ spin is responsible for a new  perpendicularly
propagating (with respect to the magnetic field direction) electromagnetic spin wave, which can
be excited by intense neutrino bursts in supernovae. The newly derived nonlinear EMHD and
Hall-MHD equations can be used for investigating the multi-dimensional linear
and nonlinear electromagnetic waves in quantum plasmas. Finally, we have
also presented theoretical and numerical studies of stimulated Raman and Brillouin scattering
instabilities of a large amplitude electromagnetic wave, and the trapping of arbitrary large
amplitude circularly polarized EM waves in a fully nonlinear electron density hole in an unmagnetized
quantum plasma. It is expected that localized nanostructures can transport electromagnetic wave
energy over nanoscales in laboratory and astrophysical dense plasmas with degenerate electrons.

The field of the nonlinear quantum plasma physics is extremely rich and vibrant
today, and it holds a great promise of providing new practical technologies. For example,
the plasma assisted carbon nanostructures and nanomaterials are the future of nanotechnologies,
so are the new radiation sources in the x-rays and gamma-rays regimes. In such circumstances,
one should have a complete understanding of the fundamentals of collective nonlinear
interactions (e.g. intense high-order harmonic generation of ultrashort laser pulses from
laser-irradiated dense plasma surfaces) in quantum plasmas. Furthermore, in magnetars,
and in the next generation intense laser-solid density plasma interaction experiments,
one would certainly have degenerate positrons, besides degenerate electrons.
The physics of dense quantum magnetoplasmas with degenerate electron-positron
pairs is expected to be quite different than what has been described in this paper.
The reason is the complex nonlinear dynamics of the electron-positron pairs, which
would have relativistic velocities in a dense magnetoplasma. Accordingly, we should develop
new theories involving relativistic kinetic and quantum relativistic magnetohydrodynamic equations
that include  quantum relativistic effects \cite{Hakim78,Shabad91,Lamata07}, electromagnetic
forces, angular momentum spin and nonlinear effects on equal footings. A detailed
analysis of such theories would provide us a guide line for understanding the
origin of localized high-energy radiation and other complex phenomena (e.g. the formation
of structures) from astrophysical settings and future laboratory experiments aiming to model
astrophysical scenarios.

{\bf Acknowledgements}

One of the authors (P K Shukla) acknowledges the benefit of useful discussions with
Academician professor Vladimir Fortov. We greately appreciate our invaluable collaborations with Lennart Stenflo, Gert Brodin, Mattias Marklund, Fernando Haas and Nitin Shukla.

This research was partially supported by the Deutsche Forschungsgemeinschaft
(Bonn, Germany) through the project SH21/3-1 of the Research Unit 1048,
and by the Swedish Research Council (VR).

\newpage

\appendix

\section{Derivation of the Vlasov equation from the Wigner equation}
We here show that the Wigner equation converges to the Vlasov equation
in the classical limit $\hbar\rightarrow 0$.
The Wigner equation reads
\begin{equation}
  \frac{\partial f}{\partial t}+{\bf v}\cdot \nabla f
  =-\frac{i e m_e^3}{(2\pi)^3\hbar^4}\int \int d^3\lambda d^3v'
  \exp\left[i \frac{m_e}{\hbar} ({\bf v}-{\bf v}')
  \cdot{\boldsymbol{\lambda}}\right]
  \left[
  \phi\left({\bf x}+\frac{\boldsymbol{\lambda}}{2},t\right)
 -\phi\left({\bf x}-\frac{\boldsymbol{\lambda}}{2},t\right)\right]f({\bf x},{\bf v}',t).
  \label{w_eq1}
\end{equation}
By the change of variables $\boldsymbol{\lambda}=\hbar \boldsymbol{\xi}/m_e$ we
have $d^3\lambda=\hbar^3d^3\xi/m_e^3$ and Eq. (\ref{w_eq1}) takes the form
\begin{equation}
  \frac{\partial f}{\partial t}+{\bf v}\cdot \nabla f
  =-\frac{i e}{(2\pi)^3\hbar}\int \int d^3\xi d^3v'
  \exp[i ({\bf v}-{\bf v}')\cdot{\boldsymbol{\xi}}]
  \left[
  \phi\left({\bf x}+\frac{\hbar\boldsymbol{\xi}}{2m_e},t\right)
 -\phi\left({\bf x}-\frac{\hbar\boldsymbol{\xi}}{2m_e},t\right)\right]f({\bf x},{\bf v}',t).
  \label{w_eq2}
\end{equation}
Assuming that $\hbar/m_e$ is small and Taylor expanding $\phi$ up to the third order
around $\bf x$,
\begin{equation}
  \phi\left({\bf x}\pm\frac{\hbar\boldsymbol{\xi}}{2m_e},t\right)
  \approx\phi({\bf x},t)\pm\frac{\hbar}{2m_e}(\boldsymbol{\xi}\cdot\nabla)\phi({\rm x},t)
  +\frac{\hbar^2}{8 m_e^2}(\boldsymbol{\xi}\cdot\nabla)^2\phi({\rm x},t)
  \pm\frac{\hbar^3}{48 m_e^3}(\boldsymbol{\xi}\cdot\nabla)^3\phi({\rm x},t),
  \label{w_eq3}
\end{equation}
we have
\begin{equation}
  \frac{\partial f}{\partial t}+{\bf v}\cdot \nabla f
  \approx-\frac{i e}{(2\pi)^3 m_e}\int \int d^3\xi d^3v'
  \exp[i ({\bf v}-{\bf v}')\cdot{\boldsymbol{\xi}}]
    \left[\boldsymbol{\xi}\cdot\nabla\phi({\rm x},t)+\frac{\hbar^2}{24 m_e^2}
(\boldsymbol{\xi}\cdot\nabla)^3\phi({\rm x},t)\right]
   f({\bf x},{\bf v}',t).
  \label{w_eq4}
\end{equation}
By the identity
\begin{equation}
\exp[i ({\bf v}-{\bf v}')\cdot{\boldsymbol{\xi}}]\boldsymbol{\xi}=
i\nabla_{{\bf v}'}\exp[i ({\bf v}-{\bf v}')\cdot{\boldsymbol{\xi}}],
\end{equation}
where $\nabla_{{\bf v}'}=\widehat{\bf x}\partial/\partial v_x'
+\widehat{\bf y}\partial/\partial v_y'
+\widehat{\bf z}\partial/\partial v_z'$, we have
\begin{equation}
  \frac{\partial f}{\partial t}+{\bf v}\cdot \nabla f
  =\frac{e}{(2\pi)^3 m_e}\int \int d^3\xi d^3v'
    \bigg\{\exp[i ({\bf v}-{\bf v}') \cdot\boldsymbol\xi]\bigg[
    (\overleftarrow{\nabla}_{{\bf v}'}\cdot\overrightarrow{\nabla})
    -\frac{\hbar^2}{24 m_e^2}(\overleftarrow{\nabla}_{{\bf v}'}\cdot\overrightarrow{\nabla})^3\bigg]
\phi({\rm x},t)\bigg\}
   f({\bf x},{\bf v}',t),
  \label{w_eq5}
\end{equation}
where the arrows indicate the direction of operation of the nabla operators.

Integration by parts in ${\bf v}'$ space, and using that $f\rightarrow 0$
for $|{\bf v}'|\rightarrow \infty$, now yields
\begin{equation}
  \frac{\partial f}{\partial t}+{\bf v}\cdot \nabla f
  =-\frac{e}{(2\pi)^3 m_e}\int \int d^3\xi d^3v'
    \exp[i ({\bf v}-{\bf v}')\cdot\boldsymbol{\xi}]\bigg\{
    \phi({\bf x},t) \bigg[(\overleftarrow{\nabla}\cdot\overrightarrow{\nabla}_{{\bf v}'})
    -\frac{\hbar^2}{24 m_e^2}(\overleftarrow{\nabla}\cdot\overrightarrow{\nabla}_{{\bf v}'})^3\bigg]
   f({\bf x},{\bf v}',t)\bigg\}.
  \label{w_eq6}
\end{equation}
Now the integration in $\boldsymbol{\xi}$ space can formally
be performed, with the result
\begin{equation}
  \frac{\partial f}{\partial t}+{\bf v}\cdot \nabla f
  =-\frac{e}{m_e}\int d^3v'
    \delta({\bf v}-{\bf v}')\bigg\{
    \phi({\bf x},t) \bigg[(\overleftarrow{\nabla}\cdot\overrightarrow{\nabla}_{{\bf v}'})
    -\frac{\hbar^2}{24 m_e^2}(\overleftarrow{\nabla}\cdot\overrightarrow{\nabla}_{{\bf v}'})^3\bigg]
   f({\bf x},{\bf v}',t)\bigg\}.
  \label{w_eq7}
\end{equation}
where the identity
\begin{equation}
  \int d^3\xi \exp[i ({\bf v}-{\bf v}')\cdot{\bf \xi}]=(2\pi)^3\delta({\bf v}-{\bf v}')
  \label{eq8}
\end{equation}
was used and where $\delta$ is Dirac's delta function.
Finally, integration over ${\bf v}'$ space yields
\begin{equation}
  \begin{split}
  &\frac{\partial f}{\partial t}+{\bf v}\cdot \nabla f
  =-\frac{e}{m_e}\bigg\{
    \phi({\bf x},t) \bigg[(\overleftarrow{\nabla}\cdot\overrightarrow{\nabla}_{{\bf v}})
    -\frac{\hbar^2}{24 m_e^2}(\overleftarrow{\nabla}\cdot\overrightarrow{\nabla}_{{\bf v}})^3\bigg]
   f({\bf x},{\bf v},t)\bigg\}.
  \end{split}
  \label{w_eq9}
\end{equation}
In the limit $\hbar\rightarrow 0$, we recover the Vlasov equation
\begin{equation}
  \frac{\partial f}{\partial t}+{\bf v}\cdot \nabla f
  =-\frac{e}{m_e}
    \nabla\phi({\rm x},t) \cdot\nabla_{{\bf v}}
   f({\bf x},{\bf v},t).
  \label{w_eq10}
\end{equation}

\section{Derivation of the dispersion relation for the Wigner-Poisson system}
We here present a derivation of the dispersion relation for electrostatic waves in a
degenerate quantum plasma, governed by the Wigner-Poisson system of equations.
The linearized Wigner-Poisson system of equations reads
\begin{equation}
  \frac{\partial f_1}{\partial t}+{\bf v}\cdot \nabla f_1
  =-\frac{i e m_e^3}{(2\pi)^3\hbar^4}\int \int d^3\lambda d^3v'
  \exp\left[i \frac{m_e}{\hbar} ({\bf v}-{\bf v}')
  \cdot{\boldsymbol{\lambda}}\right]
  \left[
  \phi_1\left({\bf x}+\frac{\boldsymbol{\lambda}}{2},t\right)
 -\phi_1\left({\bf x}-\frac{\boldsymbol{\lambda}}{2},t\right)\right]f_0({\bf v}'),
  \label{eq_B1}
\end{equation}
and
\begin{equation}
  \nabla^2\phi_1=4 \pi e \int f_1 d^3 v,
  \label{eq_B2}
\end{equation}
where $f_0$ denotes the background distribution function, and $f_1$ and $\phi_1$ denote
the perturbed distribution function and electrostatic potential, respectively. Fourier transforming
(\ref{eq_B1})-(\ref{eq_B2}) in space and Laplace transforming them in time, we obtain
\begin{equation}
  (\omega-{\bf k}\cdot{\bf v})f_1
  =\frac{e m_e^3}{(2\pi)^3\hbar^4}\int \int d^3\lambda d^3v'
  \exp\left[i \frac{m_e}{\hbar} ({\bf v}-{\bf v}')
  \cdot{\boldsymbol{\lambda}}\right]
  \left[
    e^{i {\bf k}\cdot\boldsymbol{\lambda}/2}-e^{-i {\bf k}\cdot\boldsymbol{\lambda}/2}
  \right]f_0({\bf v}')\phi_1(\omega,{\bf k}),
  \label{eq_B3}
\end{equation}
\begin{equation}
  k^2\phi_1=-4 \pi e\int f_1 d^3 v
  \label{eq_B4}
\end{equation}
Rewriting (\ref{eq_B3}) as
\begin{equation}
  \begin{split}
  &(\omega-{\bf k}\cdot {\bf v})f_1
  =\frac{i e m_e^3}{(2\pi)^3\hbar^4}\int \int d^3\lambda d^3v'
  \left\{
    \exp\left[\frac{m_e}{\hbar} ({\bf v}-{\bf v}')
  \cdot{\boldsymbol{\lambda}}+i {\bf k}\cdot\boldsymbol{\lambda}/2\right]
  \right.
  \\
  &\left.
  - \exp\left[i \frac{m_e}{\hbar} ({\bf v}-{\bf v}')
  \cdot{\boldsymbol{\lambda}}-i {\bf k}\cdot\boldsymbol{\lambda}/2\right]
  \right\}f_0({\bf v}')\phi_1(\omega,{\bf k}),
  \end{split}
\end{equation}
and performing the integration over $\boldsymbol{\lambda}$ space, we have
\begin{equation}
  (\omega-{\bf v}\cdot{\bf k})f_1
  =\frac{e m_e^3}{\hbar^4}\int d^3v'
   \left\{
    \delta\left[\frac{m_e}{\hbar} ({\bf v}-{\bf v}')+\frac{\bf k}{2}\right]
    \right.
    \left.
   -\delta\left[\frac{m_e}{\hbar} ({\bf v}-{\bf v}')-\frac{\bf k}{2}\right]
  \right\}f_0({\bf v}')\phi_1(\omega,{\bf k}),
\end{equation}
where $\delta$ is the Dirac delta function. Now, the integration can be performed over ${\bf v}'$ space,
obtaining the result
\begin{equation}
  \begin{split}
  &(\omega-{\bf k}\cdot{\bf v})f_1
  =\frac{e}{\hbar}
   \left[
     f_0\left({\bf v}+\frac{\hbar{\bf k}}{2m_e}\right)
    -f_0\left({\bf v}-\frac{\hbar{\bf k}}{2m_e}\right)
  \right]\phi_1(\omega,{\bf k}).
  \end{split}
  \label{eq_B7}
\end{equation}
Solving for $f_1$ in (\ref{eq_B7}) and inserting the result in (\ref{eq_B4}), we obtain the dispersion relation
\begin{equation}
  \begin{split}
  &1-\frac{4 \pi e^2k^2}{\hbar}
   \int \left[
     \frac{f_0\left({\bf v}+\frac{\hbar{\bf k}}{2m_e}\right)}{(-\omega+{\bf k}\cdot {\bf v})}
    -\frac{f_0\left({\bf v}-\frac{\hbar{\bf k}}{2m_e}\right)}{(-\omega+{\bf k}\cdot {\bf v})}
  \right] d^3 v=0.
  \end{split}
\end{equation}
Suitable changes of variables in the two terms now gives
\begin{equation}
  \begin{split}
  &1-\frac{4 pi e^2 k^2}{\hbar}
   \int \left[
     \frac{1}{[-\omega+{\bf k}\cdot({\bf u}-\frac{\hbar{\bf k}}{2 m_e})]}
    -\frac{1}{[-\omega+{\bf k}\cdot ({\bf u}+\frac{\hbar{\bf k}}{2 m_e})]}
  \right]f_0({\bf u}) d^3 u=0,
  \end{split}
\end{equation}
which can be rewritten as
\begin{equation}
  \begin{split}
  &1-\frac{4 \pi e^2}{m_e}
   \int
     \frac{f_0({\bf u})}{(\omega-{\bf k}\cdot{\bf u})^2-\frac{\hbar^2 k^4}{4 m_e^2}} d^3 u=0.
  \end{split}
  \label{eq_B10}
\end{equation}
The dispersion relation (\ref{eq_B10}) was also derived by Bohm and Pines \cite{Bohm53} using a series of canonical transformations of the Hamiltonian of the system; see for example equation (57) in their paper. We now choose a coordinate system such that the $x$ axis is aligned with the wave vector ${\bf k}$. Then, (\ref{eq_B10}) takes the form
\begin{equation}
  \begin{split}
  &1-\frac{4 \pi e^2}{m_e}
   \int
     \frac{f_0({\bf u})}{(\omega-k u_x)^2-\frac{\hbar^2 k^4}{4 m_e^2}} d^3 u=0.
  \end{split}
  \label{eq_B11}
\end{equation}
We next consider a dense plasma with degenerate electrons in the zero temperature limit. Then, the background distribution function takes the simple form
\begin{equation}
f_0=\left\{\begin{array}{cc}
2\left(\frac{m_e}{2\pi \hbar}\right)^3, & |{\bf u}|\leq V_{Fe} \\
0, & \mbox{elsewhere,}
\end{array}
\right.
\end{equation}
where $V_{Fe}=(2{\cal E}_{Fe}/m_e)^{1/2}$ is the speed of an electron on the Fermi surface, and
${\cal E}_{Fe}=(3\pi^2 n_0)^{2/3}\hbar^2/(2m_e)$ is the Fermi energy. The integration in (\ref{eq_B11}) can be performed over velocity space perpendicular to $u_x$, using cylindrical coordinate in $u_y$ and $u_z$, obtaining the result
\begin{equation}
  \begin{split}
  &1
  =\frac{4 \pi e^2}{m_e}
   \int
     \frac{F_0(u_x)}{(\omega-k u_x)^2-\frac{\hbar^2 k^4}{4 m_e^2}} d u_x,
  \end{split}
  \label{eq_B13}
\end{equation}
where
\begin{equation}
F_0(u_x)=\int\int f_0({\bf u})du_y du_z=2\pi\int_0^{\sqrt{V_{Fe}^2-u_x^2}} 2\left(\frac{m_e}{2\pi \hbar}\right)^3u_\perp\,du_\perp
=\left\{\begin{array}{cc}
2\pi\left(\frac{m_e}{2\pi \hbar}\right)^3(V_{Fe}^2-u_x^2), & |u_x|\leq V_{Fe} \\
0, & \mbox{elsewhere.}
\end{array}
\right.
\end{equation}
Equation (\ref{eq_B13}) can be written as
\begin{equation}
  \begin{split}
  &1
  =\frac{8 \pi^2 e^2}{m_e}\left(\frac{m_e}{2\pi \hbar}\right)^3
   \int_{-V_{Fe}}^{V_{Ve}}
     \frac{V_{Fe}^2-u_x^2}{(\omega-k u_x)^2-\frac{\hbar^2 k^4}{4 m_e^2}} d u_x
   =\frac{3\omega_{pe}^2}{4 V_{Fe}^3}
   \int_{-V_{Fe}}^{V_{Ve}}
     \frac{V_{Fe}^2-u_x^2}{(\omega-k u_x)^2-\frac{\hbar^2 k^4}{4 m_e^2}} d u_x.
  \end{split}
  \label{eq_B15}
\end{equation}
First, in the limit $\hbar k/m_e\rightarrow 0$, we have from (\ref{eq_B15})
\begin{equation}
  1+\frac{3\omega_{pe}^2}{k^2 V_{Fe}^2}\left(
  1-\frac{\omega}{2k V_{Fe}}\log\left|\frac{\omega+k V_{Fe}}{\omega-k V_{Fe}}\right|
  \right)=0,
  \label{eq_B16}
\end{equation}
where we have assumed that $\omega$ is real and $\omega/k>V_{Fe}$.

Expanding (\ref{eq_B16})
for small wavenumbers up to terms containing $k^2$, we have
\begin{equation}
  \omega^2=\omega_{pe}^2+\frac{3}{5}k^2 V_{Fe}^2.
\end{equation}
Second, for non-zero $\hbar k^2/m_e$, we have from (\ref{eq_B15}) the dispersion relation
\begin{equation}
  \begin{split}
  1+\frac{3\omega_{pe}^2}{4 k^2 V_{Fe}^2}\left\{
  2-\frac{m_e}{\hbar k V_{Fe}}\left[V_{Fe}^2-\left(\frac{\omega}{k}+\frac{\hbar k}{2 m_e}\right)^2\right]
  \log\left|
  \frac{
  \frac{\omega}{k}-V_{Fe}+\frac{\hbar k}{2 m_e}
  }{
  \frac{\omega}{k}+V_{Fe}+\frac{\hbar k}{2 m_e}
  }
  \right|
  \right.
  \\
  \left.
  +\frac{m_e}{\hbar k V_{Fe}}\left[V_{Fe}^2-\left(\frac{\omega}{k}-\frac{\hbar k}{2 m_e}\right)^2\right]
  \log\left|
  \frac{
  \frac{\omega}{k}-V_{Fe}-\frac{\hbar k}{2 m_e}
  }{
  \frac{\omega}{k}+V_{Fe}-\frac{\hbar k}{2 m_e}
  }
  \right|
  \right\}=0.
  \end{split}
  \label{eq_B18}
\end{equation}
Expanding (\ref{eq_B18}) for small wavenumbers up to terms containing $k^4$, we
obtain
\begin{equation}
  \omega^2\approx\omega_{pe}^2+\frac{3}{5}k^2 V_{Fe}^2+(1+\alpha)\frac{\hbar^2 k^4}{4 m_e^2},
\end{equation}
where $\alpha=(48/175)m_e^2 V_{Fe}^4/\hbar^2\omega_{pe}^2\approx 2.000 (n_0 a_0^3)^{1/3}$ and $a_0=\hbar^2/m_e e^2\approx 53\times 10^{-10}\,\mathrm{cm}$ is the Bohr radius. For a typical metal such as gold, which has a free electron number density of $n_0=5.9\times10^{22}\,\mathrm{cm}^{-3}$, we would have $\alpha\approx 0.4$. For the free electron density in semiconductors, which is many orders of magnitude less than in metals, $\alpha$ is much smaller and can safely be dropped compared to unity.

\end{document}